\documentstyle[epsfig]{mn}
\newif\ifAMStwofonts
\ifoldfss
  \ifCUPmtlplainloaded \else
    \NewTextAlphabet{textbfit} {cmbxti10} {}
    \NewTextAlphabet{textbfss} {cmssbx10} {}
    \NewMathAlphabet{mathbfit} {cmbxti10} {} 
    \NewMathAlphabet{mathbfss} {cmssbx10} {} 
  \fi
  \ifAMStwofonts
    \ifCUPmtlplainloaded \else
      \NewSymbolFont{upmath} {eurm10}
      \NewSymbolFont{AMSa} {msam10}
      \NewMathSymbol{\upi}     {0}{upmath}{19}
      \NewMathSymbol{\umu}     {0}{upmath}{16}
      \NewMathSymbol{\upartial}{0}{upmath}{40}
      \NewMathSymbol{\leqslant}{3}{AMSa}{36}
      \NewMathSymbol{\geqslant}{3}{AMSa}{3E}

      \let\leq=\leqslant 
       
    \fi
  \fi
\fi 

\ifnfssone
  \newmathalphabet{\mathit}
  \addtoversion{normal}{\mathit}{cmr}{m}{it}
  \addtoversion{bold}{\mathit}{cmr}{bx}{it}
  \newmathalphabet{\mathbfit} 
  \addtoversion{normal}{\mathbfit}{cmr}{bx}{it}
  \addtoversion{bold}{\mathbfit}{cmr}{bx}{it}
  \newmathalphabet{\mathbfss} 
  \addtoversion{normal}{\mathbfss}{cmss}{bx}{n}
  \addtoversion{bold}{\mathbfss}{cmss}{bx}{n}
  \ifAMStwofonts
    \ifCUPmtlplainloaded \else
      \UseAMStwoboldmath
      \makeatletter
      \new@mathgroup\upmath@group
      \define@mathgroup\mv@normal\upmath@group{eur}{m}{n}
      \define@mathgroup\mv@bold\upmath@group{eur}{b}{n}
      \edef\UPM{\hexnumber\upmath@group}
      \new@mathgroup\amsa@group
      \define@mathgroup\mv@normal\amsa@group{msa}{m}{n}
      \define@mathgroup\mv@bold\amsa@group{msa}{m}{n}
      \edef\AMSa{\hexnumber\amsa@group}
      \makeatother
      \mathchardef\upi="0\UPM19
      \mathchardef\umu="0\UPM16
      \mathchardef\upartial="0\UPM40
      \mathchardef\leqslant="3\AMSa36
      \mathchardef\geqslant="3\AMSa3E

      \let\leq=\leqslant 

    \fi
  \fi
\fi 

\ifnfsstwo
  \DeclareMathAlphabet{\mathbfit}{OT1}{cmr}{bx}{it}
  \SetMathAlphabet\mathbfit{bold}{OT1}{cmr}{bx}{it}
  \DeclareMathAlphabet{\mathbfss}{OT1}{cmss}{bx}{n}
  \SetMathAlphabet\mathbfss{bold}{OT1}{cmss}{bx}{n}
  \ifAMStwofonts
    \ifCUPmtlplainloaded \else
      \DeclareSymbolFont{UPM}{U}{eur}{m}{n}
      \SetSymbolFont{UPM}{bold}{U}{eur}{b}{n}
      \DeclareSymbolFont{AMSa}{U}{msa}{m}{n}
      \DeclareMathSymbol{\upi}{0}{UPM}{"19}
      \DeclareMathSymbol{\umu}{0}{UPM}{"16}
      \DeclareMathSymbol{\upartial}{0}{UPM}{"40}
      \DeclareMathSymbol{\leqslant}{3}{AMSa}{"36}
      \DeclareMathSymbol{\geqslant}{3}{AMSa}{"3E}

      \let\leq=\leqslant 

    \fi
  \fi
\fi 

\ifCUPmtlplainloaded \else
  \ifAMStwofonts \else 
    \def\upi{\pi}
    \def\umu{\mu}
    \def\upartial{\partial}
  \fi
\fi

\title{Self-similar collapse with cooling and heating in an expanding universe}
\author[Shuji Uchida and Tatsuo Yoshida]
       {Shuji Uchida and Tatsuo Yoshida\\
        Faculty of Science, Ibaraki
                University, 2-1-1 Bunkyo, Mito, Ibaraki, 310-8512, Japan}
\date{Accepted **** ***********.
      Received **** ***********;
      in original form **** ***********}

\pagerange{\pageref{firstpage}--\pageref{lastpage}}
\pubyear{2002}
\begin{document}
\maketitle
\label{firstpage}
\begin{abstract}
We derive self-similar solutions including cooling and heating in an Einstein 
de-Sitter universe, and investigate the effects of cooling and 
heating on the gas density and temperature distributions. We assume that the 
cooling rate has a power-law dependence on the gas density and temperature, 
$\Lambda$$\propto$$\rho^{A}T^{B}$, and the heating rate is 
$\Gamma$$\propto$$\rho T$. The values of $A$ and $B$ are chosen by requiring 
that the cooling time is proportional to the Hubble time in order to obtain 
similarity solutions. In the region where the cooling rate 
is greater than the heating rate, a cooling inflow is established, 
and the gas is compressed and heats up. Because the compression is greater 
in the inner region than in the outer region, the temperature becomes an 
increasing profile toward the center. In particular, when a large infall 
velocity is produced due to an enormous energy loss, 
the slope of the density approaches a value that depends on 
$A$, $B$, and the velocity slope, and the slope of the temperature 
approaches $-$1. On the other hand, in the region where the heating rate 
is greater than the cooling rate, the infall velocity is suppressed, 
compression of the gas is weakened, and the gas cools down. 
The slope of the density becomes shallow due to suppression of the 
contraction, and the temperature is lower than that without heating. 
The self-similar collapse presented here gives insights to 
the effects of cooling and heating on the gas distributions in galaxies 
and clusters of galaxies.   
\end{abstract}
\begin{keywords}
cosmology: theory - galaxies: clusters: general - gravitation - hydrodynamics 
- intergalactic medium - methods: analytical - radiation mechanisms: general.
\end{keywords}
\section[]{Introduction}
Clusters of galaxies contain large quantities of hot gas. 
The gas collapses under the influence of gravity, and loses 
thermal energy due to radiative cooling. Especially, the effect of 
cooling is important in the inner region where the cooling time is less than 
the Hubble time. Inside this region, it is expected that the internal 
pressure decreases, the gas cools down and cooling flows occur. 
However, from recent observations of clusters of galaxies, 
the expected cooled gas is not detected in the central region
(Peterson et al. 2001; Tamura et al. 2001).
Furthermore, X-ray observations indicate that the observed 
luminosity-temperature relation differs from the self-similar relation
(Arnaud \& Evrard 1999; Helsdon \& Ponman 2000). 
As an origin of this discrepancy, it is supported that the gas 
is significantly affected by non-gravitational heating
(Kaiser 1991; Evrard \& Henry 1991).
It is very important to investigate the influences of cooling and heating 
on the gas distributions, and it would provide an understanding of 
the thermal evolution of the intracluster medium.

For adiabatic similarity solutions in an Einstein de-Sitter universe,  
Bertschinger(1985) derived the similarity solution of a top-hat density 
perturbation, $n$=3, where $n$ is a power spectrum index, and 
showed that the density has a power-law profile against the radius, 
$\rho$$\propto$$r^{-2.25}$. Chuzhoy \& Nusser\shortcite{Chuzhoy} obtained 
the asymptotic behaviour of similarity variables of adiabatic gas, and 
found that they strongly depend on the initial power spectrum.
For $n>$$-$2, they found that the density and temperature asymptotically 
approach $\rho$$\propto$$r^{-3(n+3)/(n+5)}$ and 
$T$$\propto$$r^{-(n-1)/(n+5)}$ in the limit $r$$\rightarrow$0, respectively.  
For a self-similar solution with cooling, Abadi et al.\shortcite{Abadi} 
obtained a similarity solution of collisional gas in an Einstein de-Sitter 
universe, and used it to evaluate the ability of SPH simulations.  
In their solution, they assumed the cooling function 
$\Lambda$$\propto$$\rho^{3/2}T$. This power-law dependence is determined 
by the requirement that the cooling time must have a fixed fraction to 
the dynamical time in order to obtain the similarity solution. 
They found that radiative cooling decreases the pressure support, 
and establishes a cooling inflow. As a result, 
the profiles of fluid variables differ from those of the adiabatic solution. 
In particular, when the energy loss is large enough to produce a large 
infall velocity, they found that the density and temperature 
approach $\rho \propto r^{-2}$ and $T\propto r^{-1}$ as $r \rightarrow$0, 
respectively. 

The purpose of this study is to investigate the effects of cooling and heating 
on the gas distributions using the similarity solution. The cooling rate 
per unit volume is assumed to be given by the power-law dependences on the 
density and temperature, $\Lambda$$\propto$$\rho^{A}T^{B}$, 
where $A$ and $B$ are related to the initial power spectrum. 
The heating rate is assumed to be proportional to the gas density and 
temperature, $\Gamma$$\propto$$\rho T$. 
Of course, the similarity solution is not directly applicable to the real 
structure, and cannot include detailed physical processes. However, it is 
possible to show exact structures of flows and to give clues concerning 
the effects of cooling and heating on the gas distributions of galaxies 
and clusters of galaxies. 

The layout of the paper is as follows. In Section 2, we present our model. 
Our results are shown in Section 3. Finally, we discuss and summarize 
our results in Section 4.

\section{Model}
\subsection{Spherical collapse in an expanding universe}
We consider here that an over-dense region has a density contrast 
($\delta_{i}$) at an initial Hubble time of $t_{H,i}$, 
and that the background universe is an Einstein de-Sitter universe. 
The gravitational collapse in the flat universe becomes a similarity 
evolution because bounded shells extend to infinity. 
Fillmore \& Goldreich\shortcite{Fillmore} introduced a scale-free 
initial density contrast,
\begin{equation}
\delta_{i}=\frac{\delta M_{i}}{M_{i}}=(\frac{M_{i}}{M_{0}})^{-\epsilon}
      =(\frac{r_{i}}{R_{0}})^{-3\epsilon},
\label{eq:equation1}
\end{equation}
where $M_{i}$=$(4/3)\pi\rho_{H,i} r_{i}^{3}$ is the mass contained within the 
proper radius $r_{i}$, $\rho_{H,i}$=$1/(6 \pi G t^{2}_{H,i})$ is the 
background density at $t_{H,i}$, and $M_{0}$=$(4/3)\pi\rho_{H,i} R_{0}^{3}$ 
is a reference mass. 
Since $\delta M_{i}$ must be constant or increase with increasing radius, and 
$\delta_{i}$ has to decrease with increasing radius, $\epsilon$ is restricted 
to $0< \epsilon \leq 1$. The variance of the density fluctuation is
\begin{equation}
|\delta(x,t_{H})|^{2}\simeq k^{3}|\delta_{k}|^{2}
\simeq a(t_{H})^{2} M^{-\frac{n+3}{3}},
\label{eq:equation2}
\end{equation}
where $x$ is the comoving radius, and 
$|\delta_{k}|^{2}$ $\propto$ $a(t_{H})^{2} k^{n}$ is the power spectrum, 
where $a(t_{H})$ is an expansion factor. The relation between the index 
$n$ and the parameter $\epsilon$ is
\begin{equation}
n=3(2\epsilon-1).
\label{eq:equation3}
\end{equation}
We assume that the initial velocity of matter is pure Hubble flow, 
and that the initial gas pressure is zero. The expanding matter slows down 
with the growth of the perturbation. Eventually, it reaches a maximum 
radius and decouples from the Hubble expansion. Since 
$a(t_{H})\propto t_{H}^{2/3}$, the perturbation grows as 
$|\delta(r,t_{H})|^{2}$ $\propto$ 
$a(t_{H})^{n+5}r^{-(n+3)}$ $\propto$ $t_{H}^{2(n+5)/3}r^{-(n+3)}$, 
based on equation (\ref{eq:equation2}).
The density contrast at the turn-around radius is 
$\delta(r_{ta},t_{H})$=(3$\pi$/4)$^{2}-$1$\sim$4.55(Peebles 1980). Thus, 
the time evolution of the turn-around radius and the enclosed mass are 
\begin{equation}
r_{ta}(t_{H})\propto t_{H}^{\xi},
\label{eq:equation4}
\end{equation} 
\begin{equation}
m_{ta}(t_{H})\propto t_{H}^\frac{2}{3\epsilon},
\label{eq:equation5}
\end{equation}
where
\begin{equation}
\xi=\frac{2}{3}(1+\frac{1}{3\epsilon}).
\label{eq:equation6}
\end{equation}
After the turn-around, particles fall toward the center. The collapse of 
matter causes an increase of the central density. Since the gas particles 
are decelerated, a shock wave is formed and propagates outward.
The system undergoes self-similar evolution at a later time.
 
\subsection{Cooling and heating laws in similarity evolutions} 
In the case of the collapse of adiabatic gas, the physical scale of the 
system is only determined by the gravitational dynamics. When radiative 
cooling and heating are included, the similarity evolution is broken due 
to the existence of many physical scales. 
In order to maintain the similarity evolution, 
it is required that all of the physical scales are proportional to the 
dynamical scale.  The scale-free condition was constructed by ensuring that 
the cooling time is proportional to the dynamical 
time (Owen et al. 1998; Abadi et al. 2000).

We assume that the cooling rate per unit volume is 
$\Lambda$=$\Lambda_{0} \rho^{A} c_{s}^{2B}$. Here, $A$ and $B$ are 
free parameters, $\Lambda_{0}$ is the cooling coefficient, 
$\rho$ is the gas density, and $c_{s}$ is the sound speed.
The cooling time becomes 
\begin{equation}
t_{cool}=\frac{E}{\frac{DE}{Dt}}
=E(\frac{\Lambda}{\rho})^{-1}
=\frac{\rho^{1-A}c_{s}^{2(1-B)}}{\gamma(\gamma-1)\Lambda_{0}}, 
\label{eq:equation7}
\end{equation}
where $E$ is the specific thermal energy of the gas, and 
$\gamma=5/3$ is the adiabatic index. The Hubble time evolves as 
\begin{equation}
t_{H}=\frac{2}{3}H_{0}^{-1}(\frac{a}{a_{0}})^{\frac{3}{2}},
\label{eq:equation8}
\end{equation}
where $a_{0}$ and $H_{0}$ are the expansion factor and 
Hubble constant at the present epoch, respectively.
The sound speed and density evolve as
\begin{equation}
c_{s}=(\frac{a}{a_{0}})^{\frac{1-n}{2(3+n)}} c_{s,0},
\label{eq:equation9}
\end{equation}
\begin{equation}
\rho=(\frac{a}{a_{0}})^{-3} \rho_{0},
\label{eq:equation10}
\end{equation}
where $c_{s,0}$ and $\rho_{0}$ are 
the sound speed and the gas density at the present epoch, respectively. 
We assume that the Hubble time is proportional to the 
cooling time by the constant $\hat{t_{c}}$, 
in order to maintain scale-free conditions
(i.e., $t_{H}$=$\hat{t_{c}}t_{cool}$),
\begin{equation}
t_{H}=\hat{t_{c}}\frac{\rho^{1-A}c_{s}^{2(1-B)}}{\gamma(\gamma-1)\Lambda_{0}}. 
\label{eq:equation10-1}
\end{equation}
Substituting equations (\ref{eq:equation8})-(\ref{eq:equation10}), 
the above equation becomes 
\begin{equation}
\frac{2}{3}H_{0}^{-1}
(\frac{a}{a_{0}})^{\frac{3}{2}-3(A-1)-\frac{1-n}{3+n}(1-B)}
=\hat{t_{c}} \frac{\rho_{0}^{1-A}c_{s,0}^{2(1-B)}}
{\gamma(\gamma-1)\Lambda_{0}}.
\label{eq:equation11}
\end{equation}
The $a$-dependence term must vanish in order to keep $\hat{t_{c}}$ constant. 
Using equation (\ref{eq:equation3}), we find 
\begin{equation}
\frac{3}{2}-3(A-1)-\frac{2-3\epsilon}{3\epsilon}
(1-B)=0.
\label{eq:equation12}
\end{equation}

Figure 1 displays the parameter space of $A$ and $B$ in relation to 
$\epsilon$. The shaded areas correspond to $0< \epsilon \leq1$. 
The solid lines display the lines of $\epsilon$=1, 2/3, 1/3, and 0. 
Abadi et al.(2000) selected at the point ($A$,$B$)=(3/2,1), which 
is independent of $\epsilon$. The points of free-free emission, 
($A$,$B$)=(2,1/2), and line cooling, ($A$,$B$)=(2,$-$1/2), 
are indicated by crosses. 

Next, we assume that the heating rate per unit volume is given by 
\begin{equation}
\Gamma=\alpha \rho c^{2}_{s},
\label{eq:equation12-1}
\end{equation} 
where $\alpha$=$(t_{H,0}/t_{H})\alpha_{0}$, 
$t_{H,0}$ is the current Hubble time, and $\alpha_{0}$ is the model parameter. 
The heating time becomes 
\begin{equation}
t_{heat}=\frac{E}{\frac{DE}{Dt}}=E(\frac{\Gamma}{\rho})^{-1}
=\frac{1}{\gamma(\gamma-1)}\frac{1}{\alpha}.
\label{eq:equation12-2}
\end{equation} 
We assume that the heating time is proportional to the Hubble time 
by the constant $\hat{t_{h}}$(i.e., $t_{H}$=$\hat{t_{h}}t_{heat}$), 
\begin{equation}
\frac{2}{3}H_{0}^{-1}=\frac{\hat{t_{h}}}{\gamma(\gamma-1)\alpha_{0}}.
\label{eq:equation12-3}
\end{equation}

\begin{figure}
   \begin{center}
   \includegraphics[width=8cm,height=10.0cm,clip]{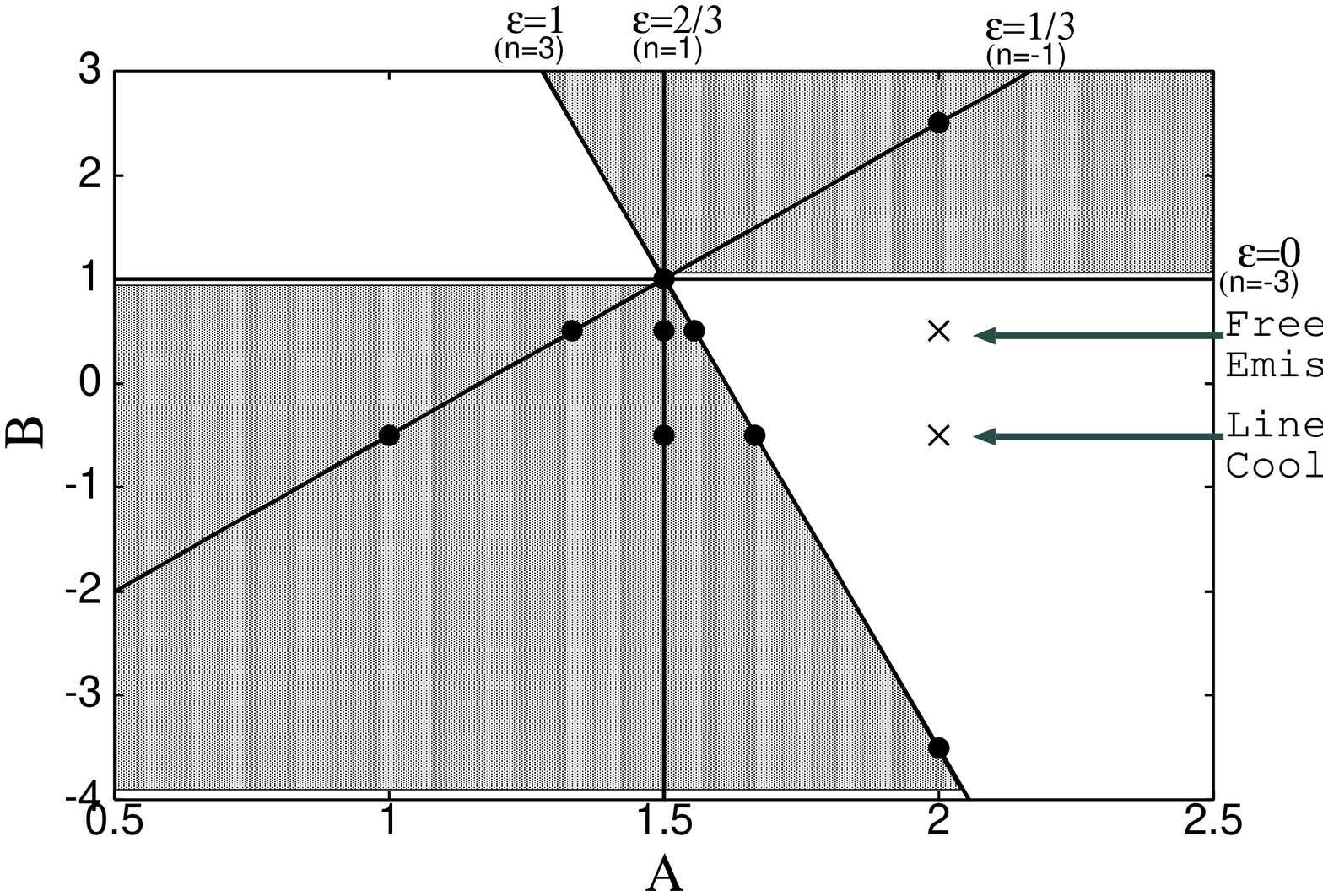}
   \caption{Relation between ($A$, $B$) and $\epsilon$. 
The shaded areas indicate the inside of the range 
$0< \epsilon \leq1$. The solid lines are constant values of 
$\epsilon$=1, 2/3, 1/3, and 0. The closed circles represent the 
parameter positions that we calculated in this paper. The points 
of free-free emission and line emission are indicated by crosses.
}
\end{center}
\label{fig:figure1}
\end{figure}%

\subsection{Basic equations of collisional gas}
Neglecting the effect of dark matter, the basic equations 
with additional terms of cooling and heating are as follows:
\begin{equation}
\frac{\partial{m}}{\partial r}=4\pi r^{2}\rho,
\label{eq:equation14}
\end{equation}
\begin{equation}
\frac{\partial\rho}{\partial t}+\frac{1}{r^{2}}\frac{\partial}
{\partial r}(r^{2}\rho v)=0,
\label{eq:equation15}
\end{equation}
\begin{equation}
\frac{\partial v}{\partial t}+v\frac{\partial v}{\partial r}=
-\frac{\partial}{\gamma \rho \partial r}
(\rho c_{s}^{2\setlength{\oddsidemargin}{-0.1in}})-\frac{Gm}{r^{2}},
\label{eq:equation16}
\end{equation}
\begin{equation}
\frac{1}{\gamma-1}\frac{dc_{s}^{2}}{dt}-\frac{c_{s}^{2}}{\rho}\frac{d\rho}{dt}
=- \gamma \Lambda_{0} \rho^{A-1} c_{s}^{2B}+\gamma \alpha c_{s}^{2},
\label{eq:equation17}
\end{equation}
where $v$ is the velocity, $G$ is the gravitational constant, and 
$m$ is the mass of the gas. 

To derive self-similar solutions, we introduce the similarity variable $X$
and similarity functions, which are given by the following forms:
\begin{equation}
X=\frac{r}{r_{ta}},
\label{eq:equation18}
\end{equation}
\begin{equation}
\rho=\rho_{H}D,
\label{eq:equation20}
\end{equation}
\begin{equation}
v=\frac{r_{ta}}{t_{H}}V,
\label{eq:equation19}
\end{equation}
\begin{equation}
c_{s}=\frac{r_{ta}}{t_{H}}C,
\label{eq:equation21}
\end{equation}
\begin{equation}
m=\frac{4}{3}\pi\rho_{H}r^{3}_{ta}M,
\label{eq:equation22}
\end{equation}
\begin{equation}
\alpha=\frac{\mathcal{H}}{t_{H}},
\label{eq:equation24}
\end{equation}
where $\rho_{H}$=$1/(6 \pi G t^{2}_{H})$ is the mean density of the 
universe, and ${\mathcal{H}}$ is the same as $\alpha_{0}t_{H,0}$ or 
$3H_{0}{\hat{t_{h}}}t_{H,0}/2\gamma(\gamma-1)$ from equation 
(\ref{eq:equation12-3}). The density and the sound speed are expressed as 
\begin{equation}
\rho=\rho_{0}\frac{\rho_{H}}{\rho_{H,0}},
\label{eq:equation24-1}
\end{equation}
\begin{equation}
c_{s}=c_{s,0}(\frac{r_{ta}}{t_{H}})/(\frac{r_{ta,0}}{t_{H,0}}),
\label{eq:equation24-2}
\end{equation}
where $\rho_{H,0}$ and $r_{ta,0}$ are the background density and the 
turn-around radius at the present epoch, respectively. 
We substitute equations (\ref{eq:equation24-1}) and (\ref{eq:equation24-2}) 
for (\ref{eq:equation10-1}), and define $\Lambda_{0}$ as   
\begin{eqnarray}
\Lambda_{0}&=&\frac{\hat{t_{c}}(\frac{\rho_{0}}{\rho_{H,0}})^{1-A}
(\frac{c_{s,0}t_{H,0}}{r_{ta,0}})^{2(1-B)}}{\gamma(\gamma-1)}
\rho_{H}^{1-A} r_{ta}^{2(1-B)}t_{H}^{2B-3},\nonumber \\
&=&\frac{K_{0}}{\gamma(\gamma-1)}\rho_{H}^{1-A} r_{ta}^{2(1-B)}t_{H}^{2B-3}.
\label{eq:equation25}
\end{eqnarray}
Substituting equations (\ref{eq:equation18})-(\ref{eq:equation24}), and 
(\ref{eq:equation25}) for 
(\ref{eq:equation14})-(\ref{eq:equation17}), the fluid equations become,
\begin{equation}
\frac{dM}{dX}=3X^{2}D,
\label{eq:equation26}
\end{equation}
\begin{eqnarray}
(V-\xi X)\frac{dD}{dX}+D\frac{dV}{dX}+2D(\frac{V}{X}-1)=0,
\label{eq:equation27}
\end{eqnarray}
\begin{eqnarray}
(V-\xi X)\frac{dV}{dX}-(1-\xi)V 
=-\frac{1}{\gamma D}\frac{d}{dX}(DC^{2})-\frac{2M}{9X^{2}}, 
\label{eq:equation28}
\end{eqnarray}
\begin{eqnarray}
\frac{2}{\gamma-1}[{(\xi-1)+\frac{(V-\xi X)}{C}\frac{dC}{dX}}]+2
-\frac{(V-\xi X)}{D}\frac{dD}{dX} \nonumber \\
=-\frac{K_{0}}{(\gamma-1)}D^{A-1}C^{2B-2}
+\gamma {\mathcal{H}}.
\label{eq:equation29}
\end{eqnarray}
The non-dimensional mass $M$ is derived by another 
integration(Bertschinger 1985).
The equation of continuity is rewritten as follows:
\begin{eqnarray}
\frac{\partial{m}}{\partial t}+4\pi r^{2}\rho v=0.
\label{eq:equation29-1}
\end{eqnarray}
Substituting equations (\ref{eq:equation18})-(\ref{eq:equation22}) for 
(\ref{eq:equation14}) and (\ref{eq:equation29-1}), the dimensionless mass is 
\begin{eqnarray}
M=\frac{-3}{3\xi-2}X^{2}D(V-\xi X).
\label{eq:equation29-2}
\end{eqnarray}

\subsection{Numerical integration and boundary conditions}
\label{subsec:numerical}
\begin{figure*}
\resizebox{8cm}{!}{\includegraphics{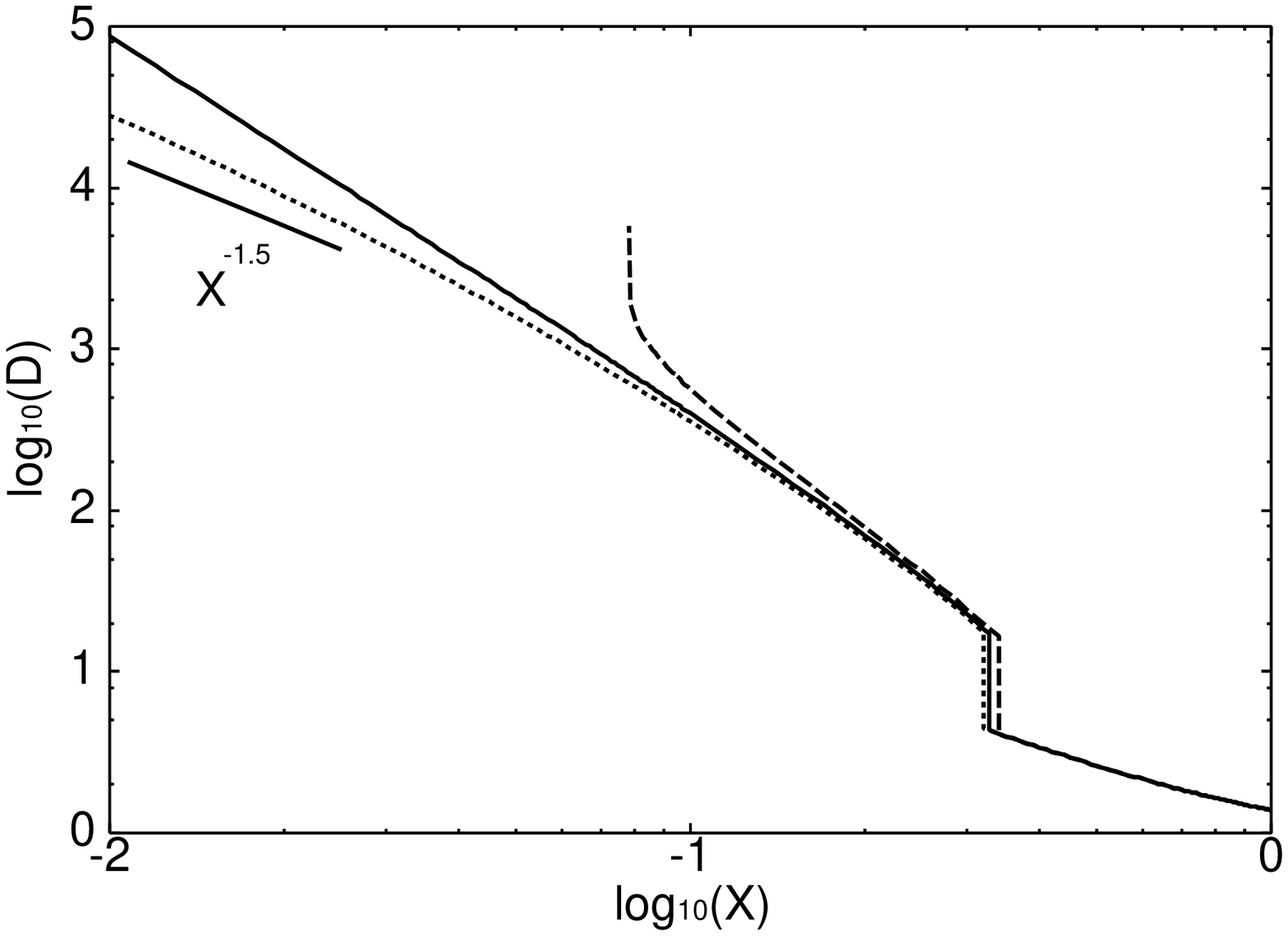}}%
\resizebox{8cm}{!}{\includegraphics{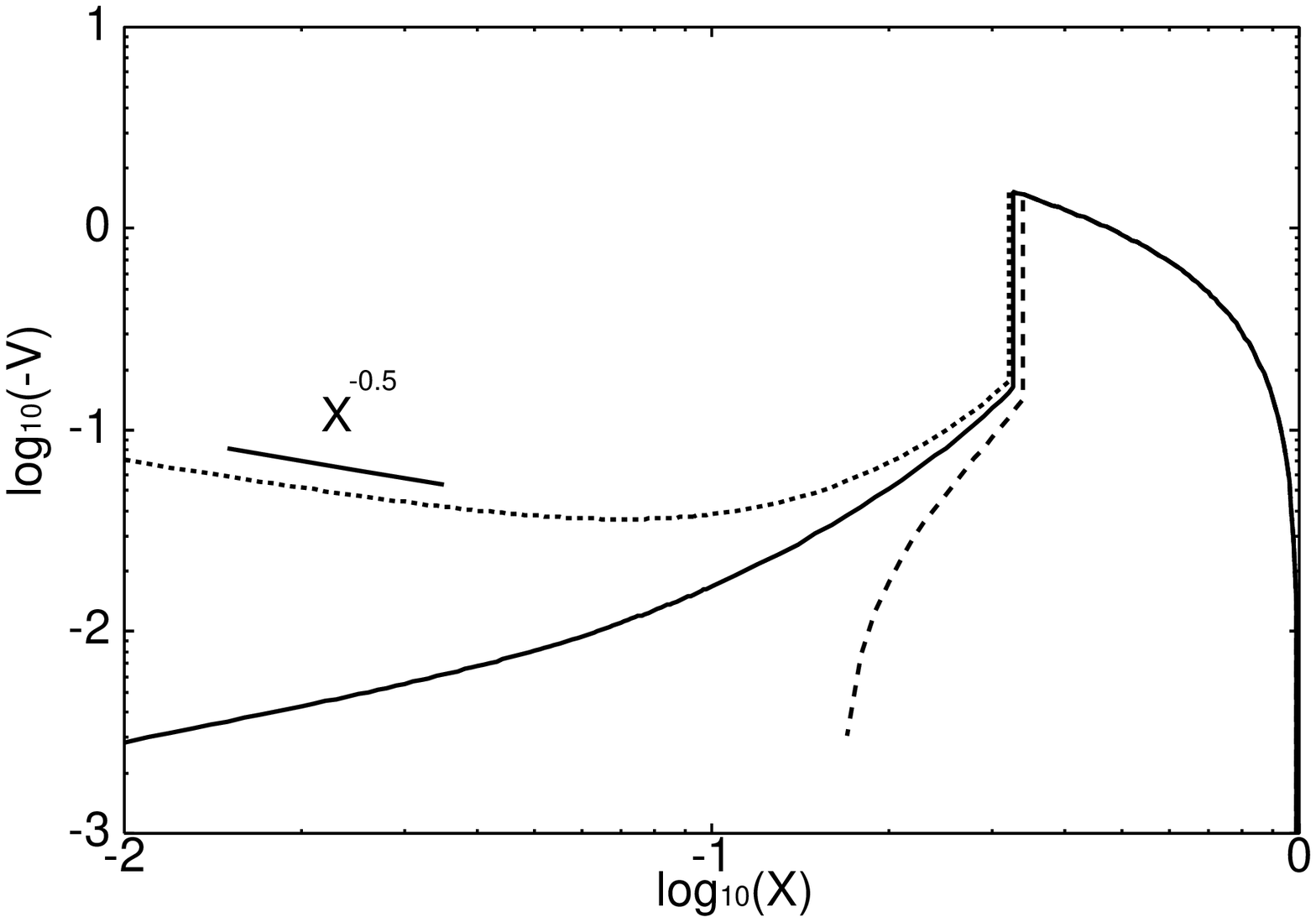}}\\
\resizebox{8cm}{!}{\includegraphics{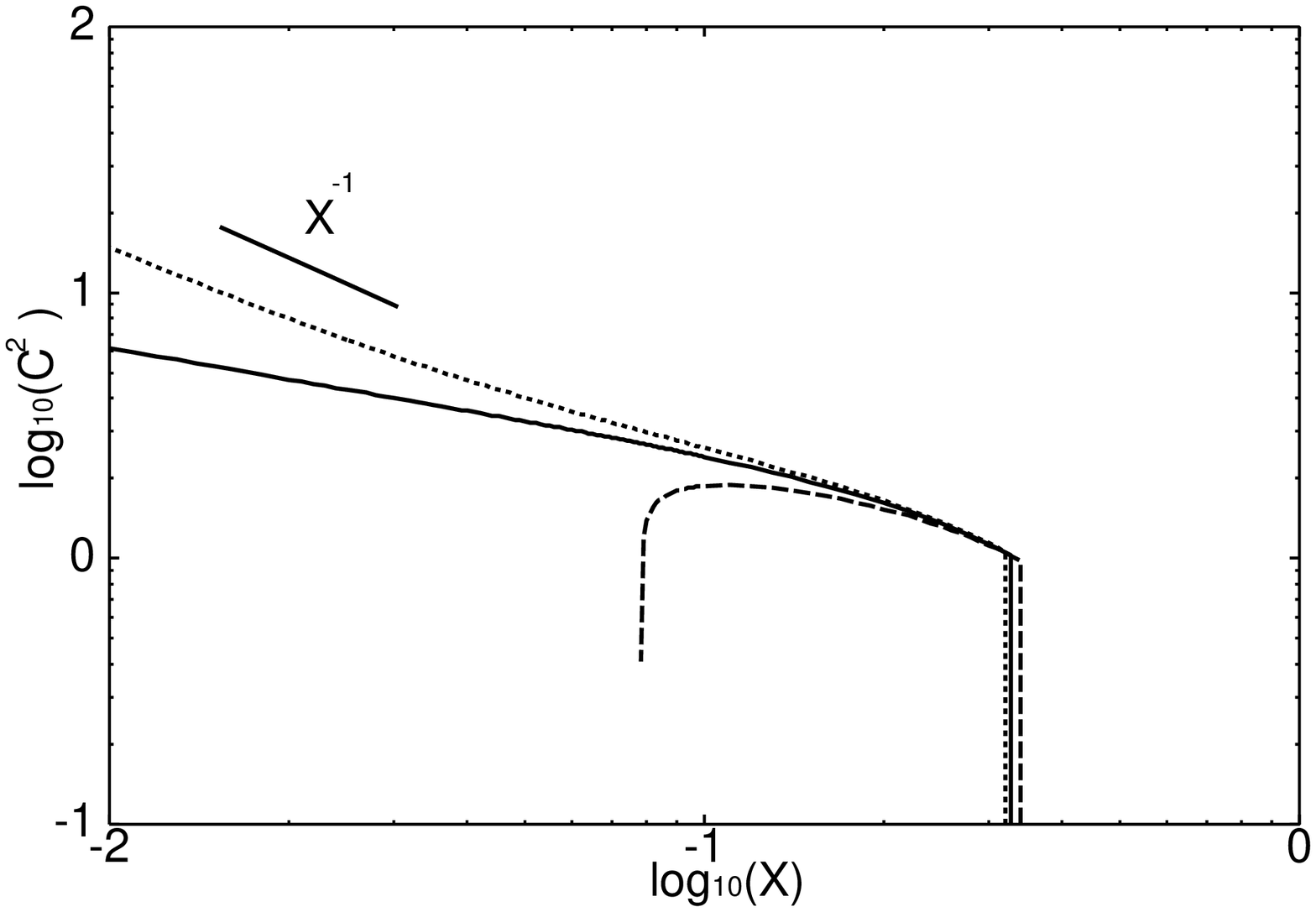}}%
\resizebox{8cm}{!}{\includegraphics{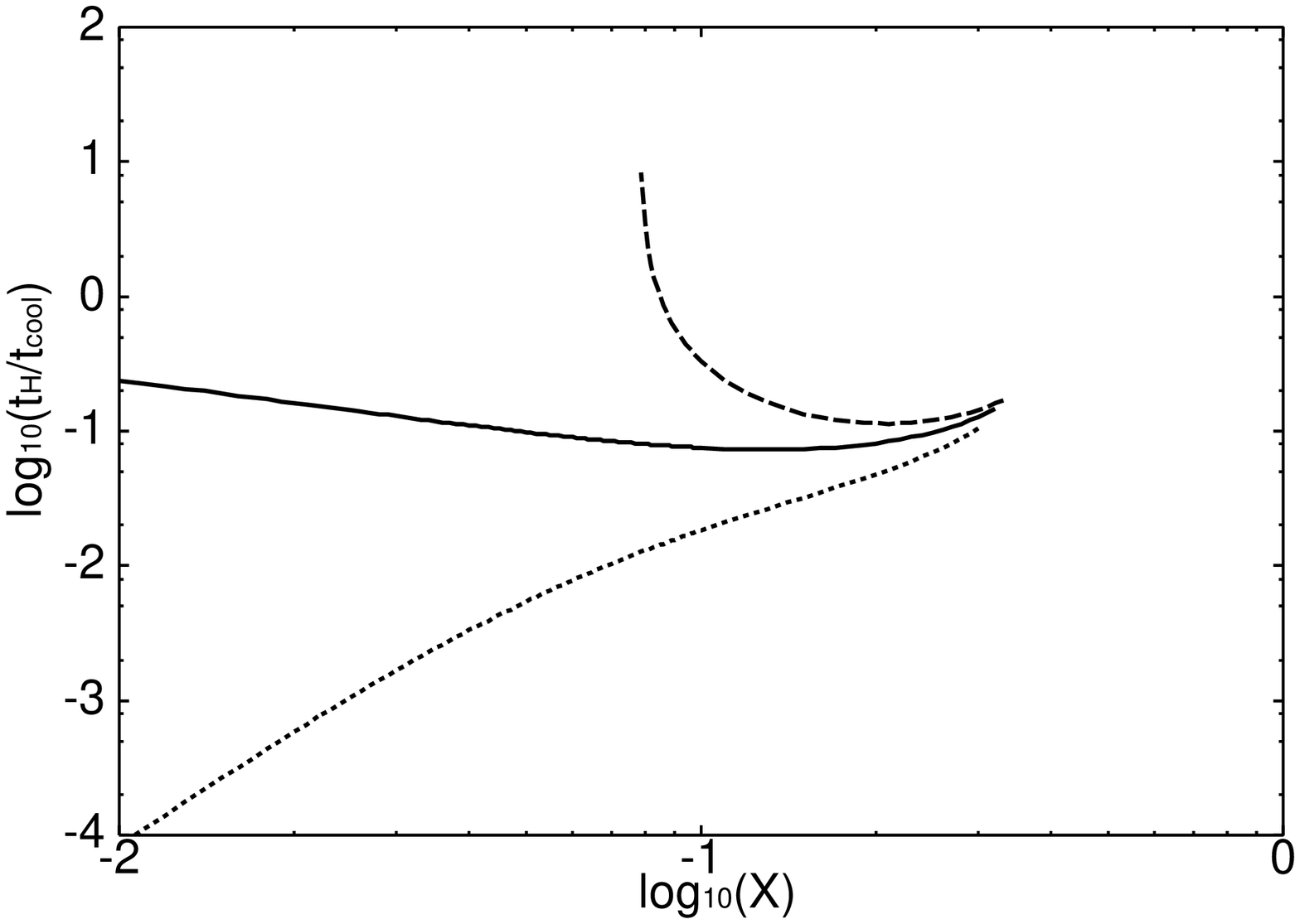}}\\
\ \\
\caption{Three types of solutions with $\epsilon$=1 ($n$=3), 
${\mathcal{H}}$=0, $A$=2, $B$=$-$7/2, and $K_{0}$=0.01. 
Each line shows the resultant profiles of the 'stagnation' solution 
(dashed line), 'eigensolution' (solid line), 
and 'adiabatic' solution (dotted line), respectively.   
The shock radii are $X_{s}$=0.338976, 0.326595, and 0.3, respectively.   
Each panel represents radial profiles of the density (top left), 
the proper velocity (top right), 
the square of the sound speed corresponding to the 
temperature (bottom left), and the ratio of the Hubble time to the 
cooling time (bottom right), respectively.}
\label{fig:figure2}
\end{figure*}%
Assuming that the gas pressure is zero, 
the physical variables of a gas shell are given by the parameter 
$\theta$(Peebles 1980):
\begin{equation}
X=\frac{r}{r_{ta}}=\rm{sin}^{2}(\frac{\theta}{2})
(\frac{\theta-\rm{sin}\theta}{\pi})^{-\xi},
\label{eq:equation30}
\end{equation}
\begin{equation}
D=\frac{9}{2}\frac{(\theta-\rm{sin}\theta)^{2}}{(1-\rm{cos}\theta)^{3}
(1+3\epsilon\chi)},
\label{eq:equation31}
\end{equation}
\begin{equation}
V=X\frac{\rm{sin}\theta(\theta-\rm{sin}\theta)}
{(1-\rm{cos}\theta)^{2}},
\label{eq:equation32}
\end{equation}
\begin{equation}
C=0,
\label{eq:equation32-1}
\end{equation}
\begin{equation}
M=\frac{9}{2}\frac{(\theta-\sin \theta)^{2}}{(1-\cos \theta)^{3}}X^{3},
\label{eq:equation33}
\end{equation}
with $\chi$=$1-3V/2X$. 
The outer boundary condition is defined at the turn-around radius, 
$\theta$=$\pi$. The collisional gas shell falls toward the center after 
the turn around, and the central gas density increases. Eventually, 
adiabatic compression occurs, and a shock propagates outward.
The shock radius appears at a fixed fraction of the 
turn-around radius. The shock jump conditions are written as
\begin{equation}
D_{2}=\frac{\gamma+1}{\gamma-1}D_{1},
\label{eq:equation34}
\end{equation}
\begin{equation}
V_{2}=\xi X_{s}+(V_{1}-\xi X_{s})\frac{\gamma-1}{\gamma+1},
\label{eq:equation35}
\end{equation}
\begin{equation}
C_{2}^{2}=\frac{2\gamma(\gamma-1)}{(\gamma+1)^{2}}(V_{1}-\xi X_{s})^{2},
\label{eq:equation36}
\end{equation}
where subscripts 1 and 2 represent the pre- and post-shock quantities, and 
$X_{s}$ is the non-dimensional shock radius. We need to determine the shock 
position. Bertschinger(1989) and Abadi et al.(2000) showed three types of 
solutions by changing the shock radius. Figure 2 represents the three 
possible solutions of the 'stagnation' solution (dashed line), 
'eigensolution' (solid line), and 'adiabatic' solution (dotted line) 
for $\epsilon$=1($n$=3), ${\mathcal{H}}$=0, 
$A$=2, $B$=$-$7/2, and $K_{0}$=0.01. Each shock radius 
is $X_{s}$=0.338976, 0.326595, and 0.3, respectively. The ratio of the 
Hubble time to the cooling time at each radius is shown in Fig. 2, 
which is defined by 
\begin{equation}
\frac{t_{H}}{t_{cool}}=K_{0}D^{A-1}C^{2B-2}.
\label{eq:equation38}
\end{equation}

The first type is the 'stagnation' solution (dashed line), which 
diverges at a small radius. 
We found that the similarity solution with $X_{s}$=0.338976, which 
corresponds to the shock position of Bertschinger's adiabatic 
solution (Bertschinger 1985), becomes the 'stagnation' solution. 
We analytically explain that the divergence at the small radius occurs 
due to cooling. Because the profiles of the fluid variables are not 
influenced by cooling in the outer region, 
we assume that the velocity approaches $V$=$V_{0}X$, 
which is the same as that of the adiabatic solution (Chuzhoy \& Nusser 2000).  
The density and temperature can be expressed from equations 
(\ref{eq:equation27}) and (\ref{eq:equation29}) as follows:
\begin{equation}
D=D_{0}X^{s},\nonumber\\
C=C_{0}X^{t},\nonumber\\
\label{eq:equation39}
\end{equation}
where 
\begin{equation}
s=\frac{1}{V_{0}-\xi}(2-3V_{0}),
\label{eq:equation40}
\end{equation}
\begin{eqnarray}
t=\frac{1}{V_{0}-\xi}[-(\xi-1)
-\frac{K_{0}}{2}D^{A-1}C^{2B-2}
-\frac{3}{2}(\gamma-1)V_{0} \nonumber \\
+\frac{\gamma(\gamma-1)}{2}{\mathcal{H}}].
\label{eq:equation41}
\end{eqnarray}
As the gas relaxes to a hydrostatic equilibrium in order to vanish the 
velocity as $X\rightarrow0$, 2$t-s$=2 is satisfied by equation 
(\ref{eq:equation28}), in which case $V_{0}$ is 
\begin{eqnarray}
V_{0}=\frac{1}{3\gamma-4}[-K_{0}D^{A-1}C^{2B-2}
+\gamma(\gamma-1)\mathcal{H}]. 
\label{eq:equation42}
\end{eqnarray}
The first and second terms on the right-hand side  
represent the cooling and heating terms, respectively.
We consider the behavior of the cooling term in the inner region in 
the case of $\mathcal{H}$=0. 
Assuming that the cooling term converges to zero at the center, 
$D^{A-1}C^{2B-2}$ becomes 
\begin{eqnarray}
D^{A-1}C^{2B-2}\propto X^{\kappa},
\label{eq:equation44}
\end{eqnarray}
where
\begin{eqnarray}
\kappa=-\frac{2}{\xi}(A-1)+\frac{\xi-1}{\xi}(2B-2)>0.
\label{eq:equation45}
\end{eqnarray}
However, the parameters $A$ and $B$, which satisfy $\kappa>$0, are not 
in the range of $0< \epsilon \leq1$. Thus, $\kappa<0$ is satisfied 
and the first term on the right-hand side of equation (\ref{eq:equation42}) 
diverges, resulting in a steeper drop of the temperature.
That is to say, a cooling catastrophe occurs in the inner region. 
 
The second type is an 'adiabatic' solution (dotted line). 
It is a free-fall solution to a point mass.  
Thus, the density and velocity asymptotically approach 
$D$$\propto$$X$$^{-3/2}$ and $V$$\propto$$X$$^{-1/2}$, respectively. 
Because the cooling and heating times are longer than the flow time, 
they are negligible for $X$$\rightarrow$0. 
For the law conservation of energy, the sound speed approaches 
$C$$\propto$$X$$^{-1/2}$ as $X\rightarrow$0. 

The third type of solution is the 'eigensolution' (solid line), 
which is a marginal case of 'stagnation' solutions. That is to say, 
the 'eigensolution' satisfies the condition $t_{flow}$=$t_{cool}$ at 
the origin. A divergence of the physical variables does not occur, 
and the flow extends to $X$=0. The 'eigensolution' is the only solution where 
the mass is zero at the origin. We only derive the 'eigensolution' below. 

\section{Results}
We derive the similarity solutions of various cases of ($A$,$B$). 
The first case is ($A$,$B$)=(3/2,1), which is the same as Abadi et al.(2000).
The others are focused on the free-free and line emissions. 
As we can see from Fig. 1, the free-free and line emissions are 
out of the range $0< \epsilon \leq1$.
Thus, we fix one parameter to $B$=1/2, $-$1/2, and $A$=2, 
in which case the other parameter, $A$ or $B$, is obtained from equation 
(\ref{eq:equation12}). 

\subsection{Adiabatic solution}
\begin{figure}
\resizebox{8cm}{!}{\includegraphics{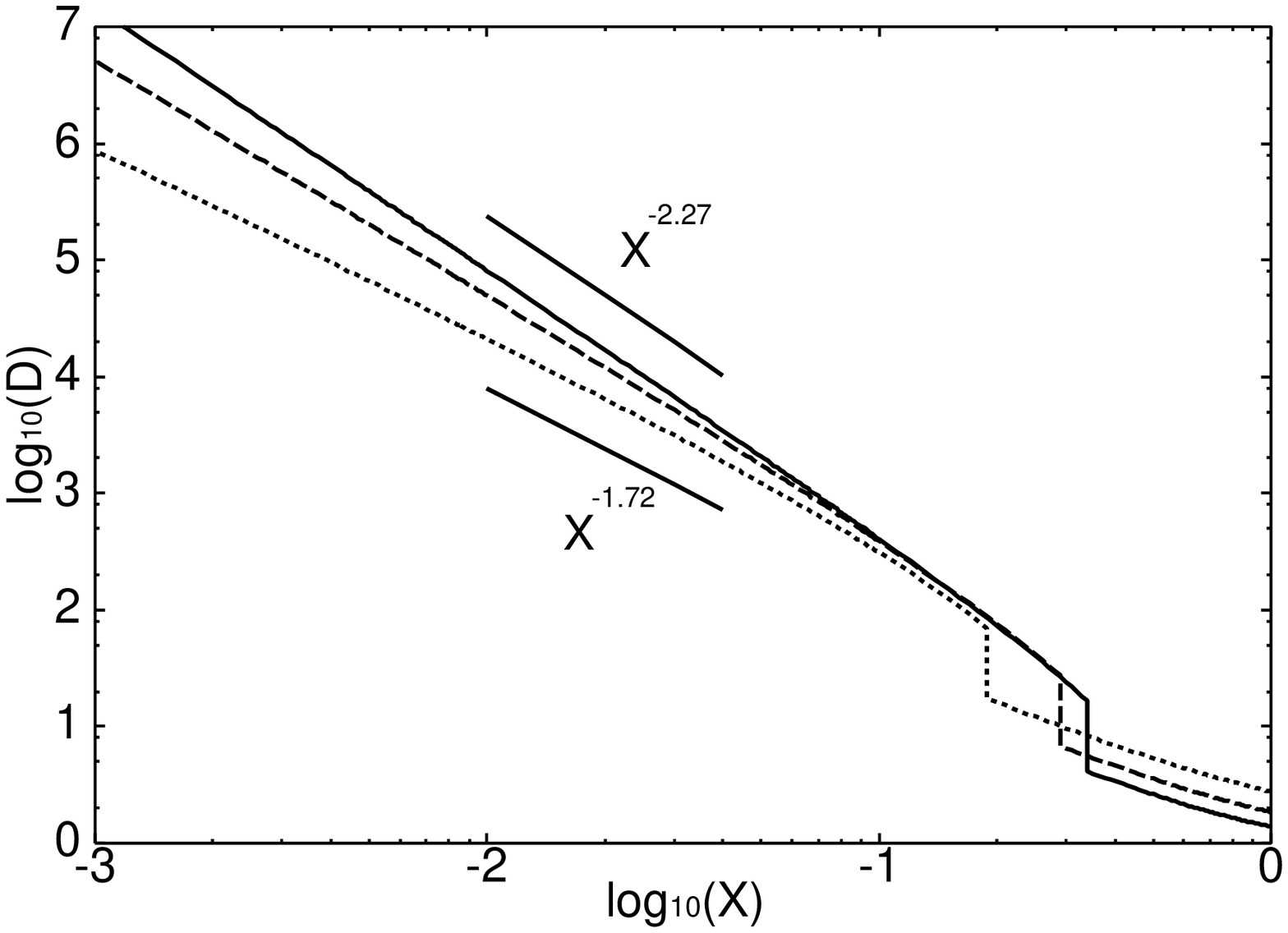}}\\
\resizebox{8cm}{!}{\includegraphics{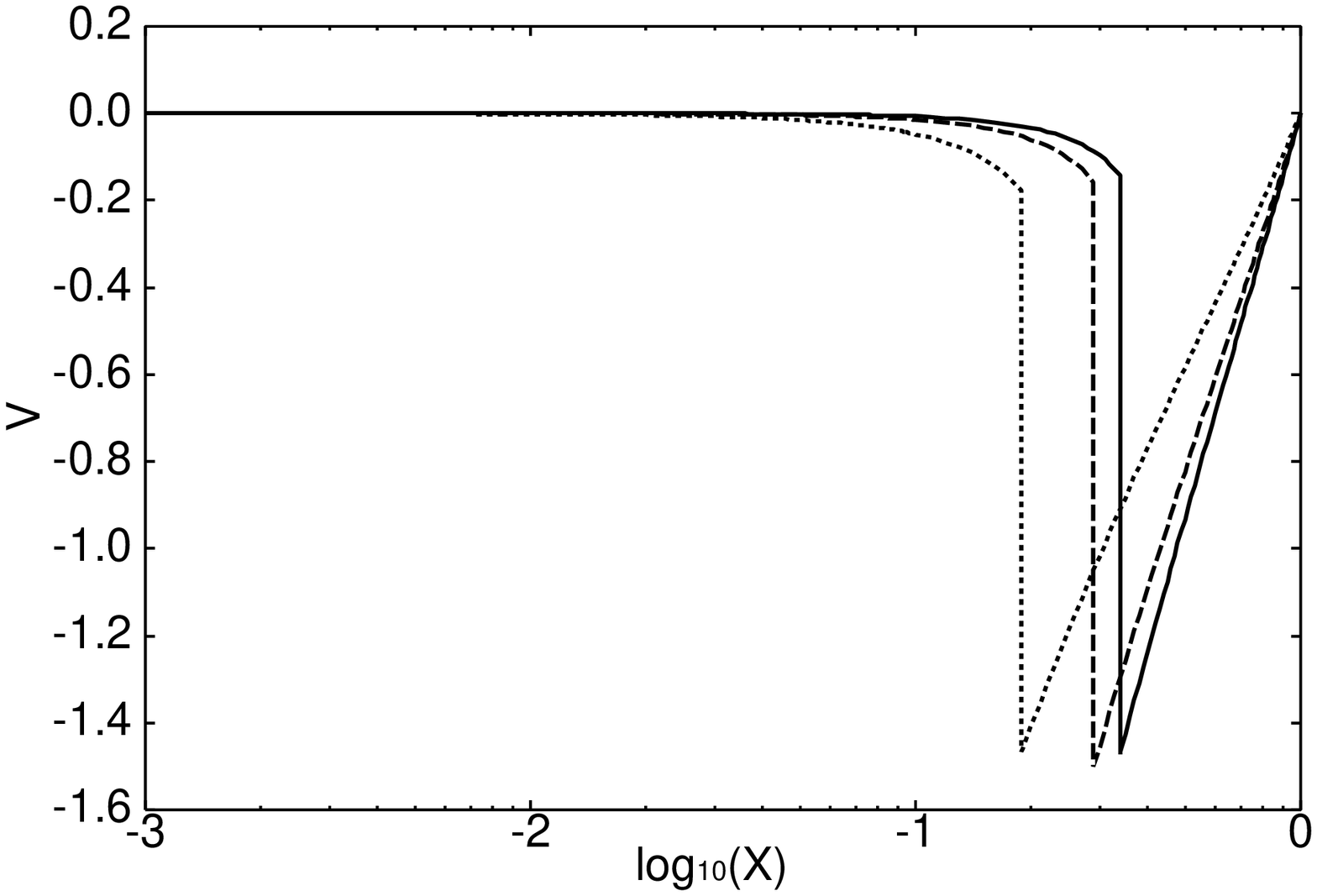}}\\
\resizebox{8cm}{!}{\includegraphics{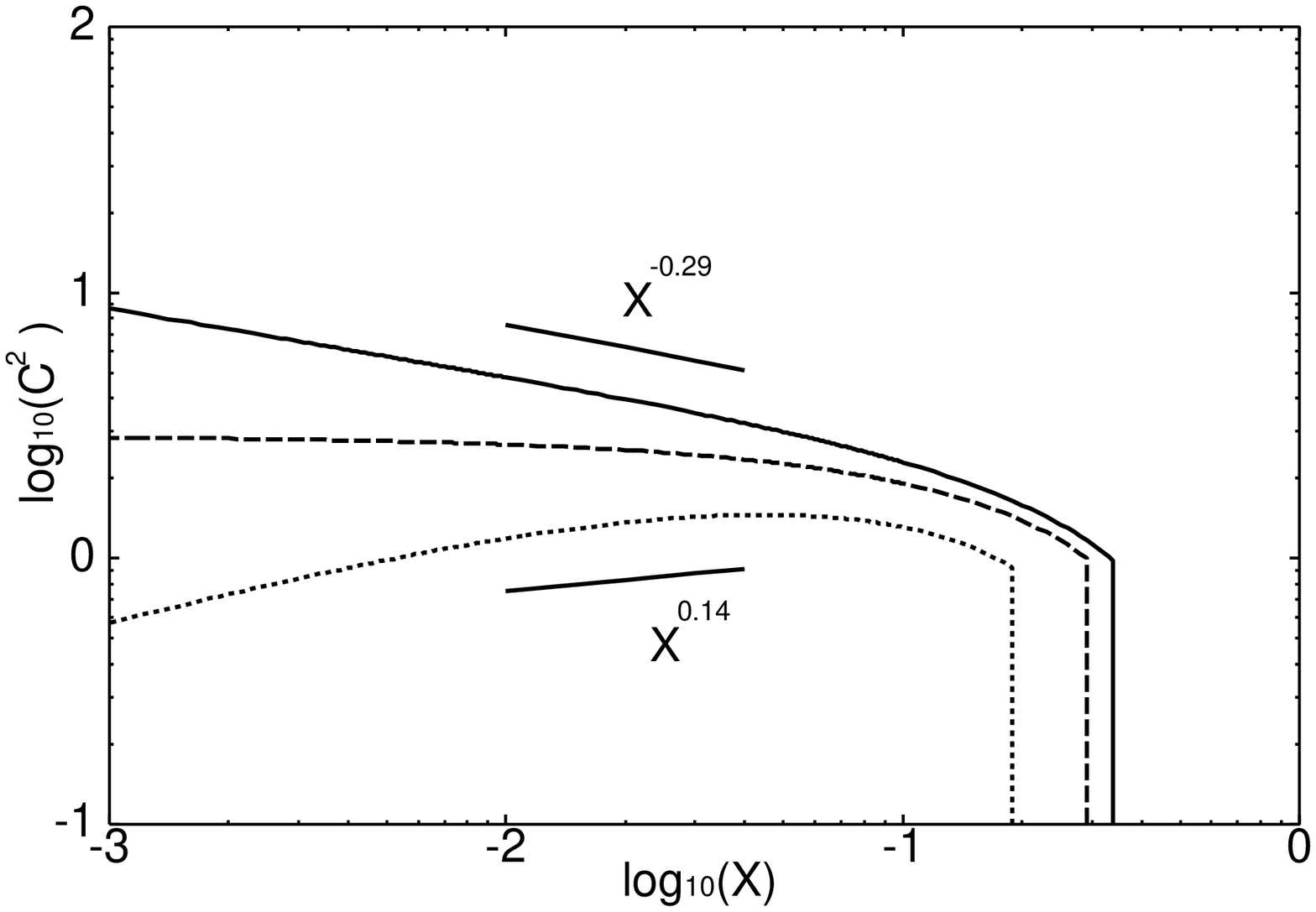}}%
\ \\
\caption{Similarity solutions without cooling and heating. 
Each line indicates the results of 
$\epsilon$=1 (solid line), $\epsilon$=2/3 (dashed line), and 
$\epsilon$=1/3 (dotted line), respectively.   
Each panel represents radial profiles of the density (top panel), 
the proper velocity (middle panel), and 
the square of the sound speed corresponding to the temperature (bottom panel), 
respectively.}
\label{fig:figure3}
\end{figure}%
\begin{table}
 \centering
 \begin{center}
  \begin{tabular}{cccc}\\
  \hline
   $\epsilon$ & $X_{s}$ & $u$ & $v$ \\
\hline 
 \hline
  1.0& 0.338976 & $-$2.27 & $-$0.29 \\
  2/3& 0.289976 & $-$2.05 & $-$0.09 \\
  1/3& 0.188952 & $-$1.72 & $+$0.14 \\  
  \hline
  \hline
\end{tabular}
\caption{Properties of similarity solutions without cooling and heating. 
$X_{s}$ is the non-dimensional shock radius, $u$ and $v$ are the gradient 
of the density and the square of the sound speed corresponding to the 
temperature which are fitted from $X$=0.01 to 0.04.}
\label{tab:Table1}
\end{center}
\end{table}%
We first represent the solutions without cooling and heating for a comparison.
Figure 3 shows the solutions for $\epsilon$=1 (solid line), 
$\epsilon$=2/3 (dashed line), and $\epsilon$=1/3 (dotted line), respectively. 
Each shock radius is $X_{s}$=0.338976 for $\epsilon$=1, 
0.289976 for $\epsilon$=2/3, and 0.188952 for $\epsilon$=1/3, respectively. 
$\epsilon$=1/3($n$=$-$1) is the closest to the slope of the CDM power spectrum 
on cluster scales (Tadros et al.1998).  
Outside of the shock radius, the gas falls freely to the center. 
After passing through the shock front, the gas velocity becomes 
nearly zero, and the gas is approximately in a hydrostatic equilibrium. 
Chuzhoy \& Nusser \shortcite{Chuzhoy} analytically obtained 
asymptotic slopes of the similarity variables, and 
found $D$$\propto$$X^{-3(n+3)/(n+5)}$ and $T$$\propto$$X^{-(n-1)/(n+5)}$ 
as $X$$\rightarrow$0 in the case of $\epsilon$$>$1/6($n$$>$$-$2). 
For $\epsilon$=1, 2/3, and 1/3, 
the density becomes $\propto$$X^{-2.25}$ for $\epsilon$=1, 
$\propto$$X^{-2}$ for $\epsilon$=2/3, and $\propto$$X^{-1.5}$ for 
$\epsilon$=1/3, and 
the temperature becomes $\propto$$X$$^{-0.25}$ for $\epsilon$=1, 
$\propto$$X^{0}$ for 
$\epsilon$=2/3, and $\propto$$X^{0.5}$ for $\epsilon$=1/3, respectively.   
From the results of a numerical calculation, 
we confirmed that their slopes do not match the analytic slopes  
because the velocity is not completely zero. 
Table 1 gives the shock radius $X_{s}$, as well as the slopes of density 
$u$ and temperature $v$, which were fitted from $X$=0.01 to 0.04. 
In the case of $\epsilon$=1/3,
the temperature shows a decreasing profile toward the center. 
The reason for this is the strength of the shock wave. 
Because the shock velocity is $v_{s}$=
$dr/dt\propto$$t^{\xi-1}$, the shock wave becomes strong with time when 
$\xi>$1.0 ($\epsilon<$2/3), resulting in a decreasing temperature profile 
toward the center. 

\subsection{Case of A=3/2, B=1}
\begin{figure*}
\resizebox{8cm}{!}{\includegraphics{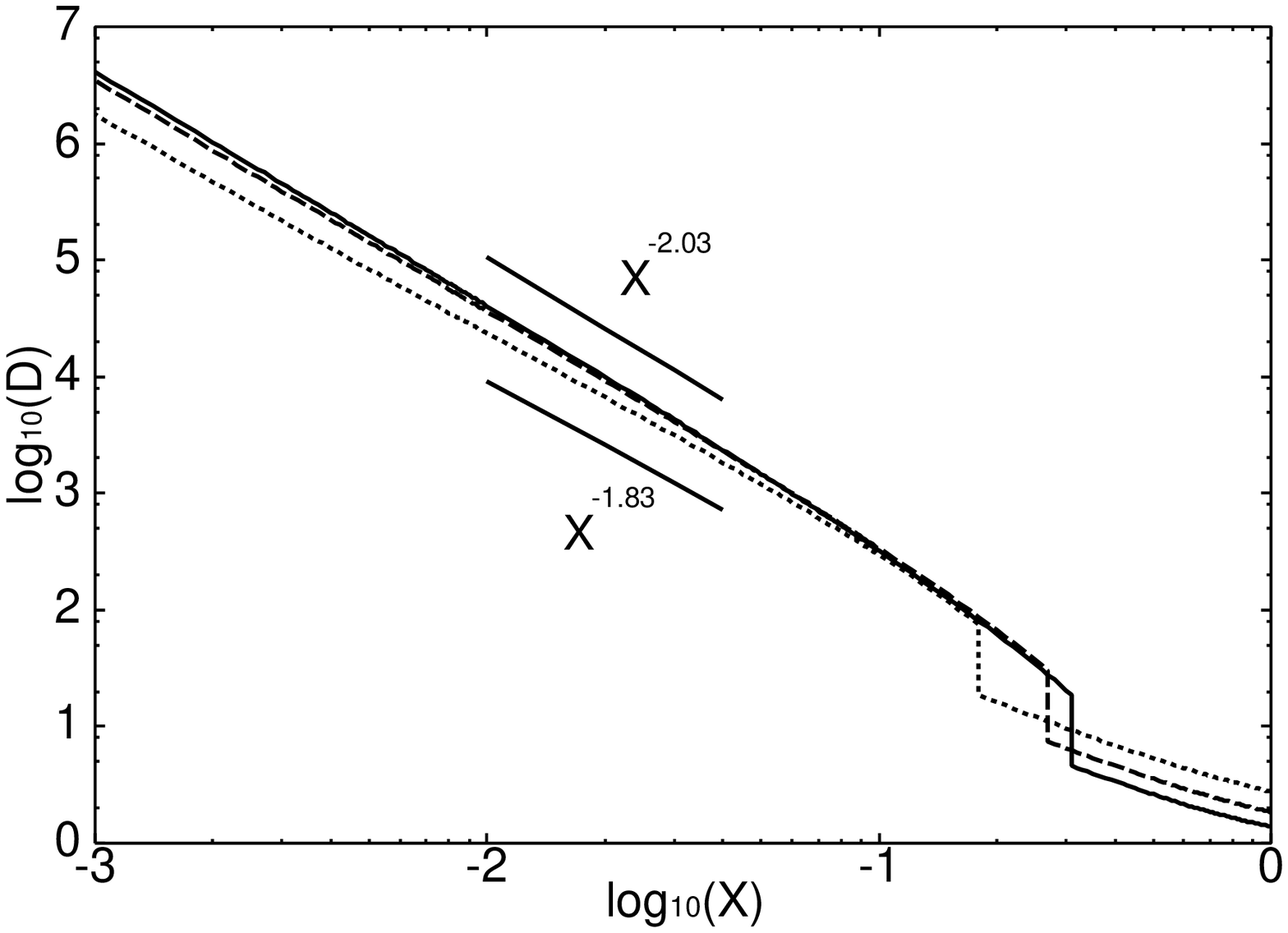}}%
\resizebox{8cm}{!}{\includegraphics{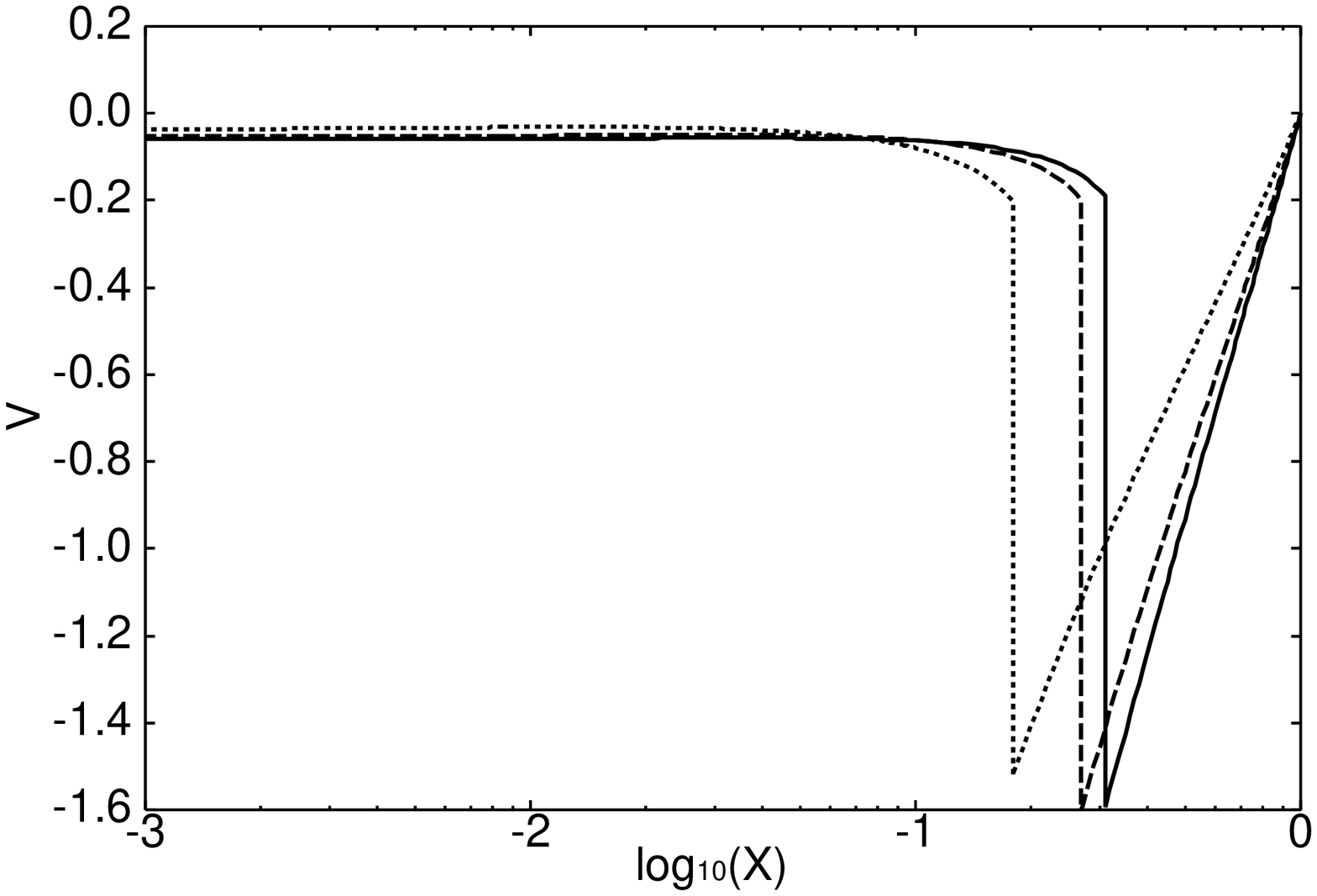}}\\
\resizebox{8cm}{!}{\includegraphics{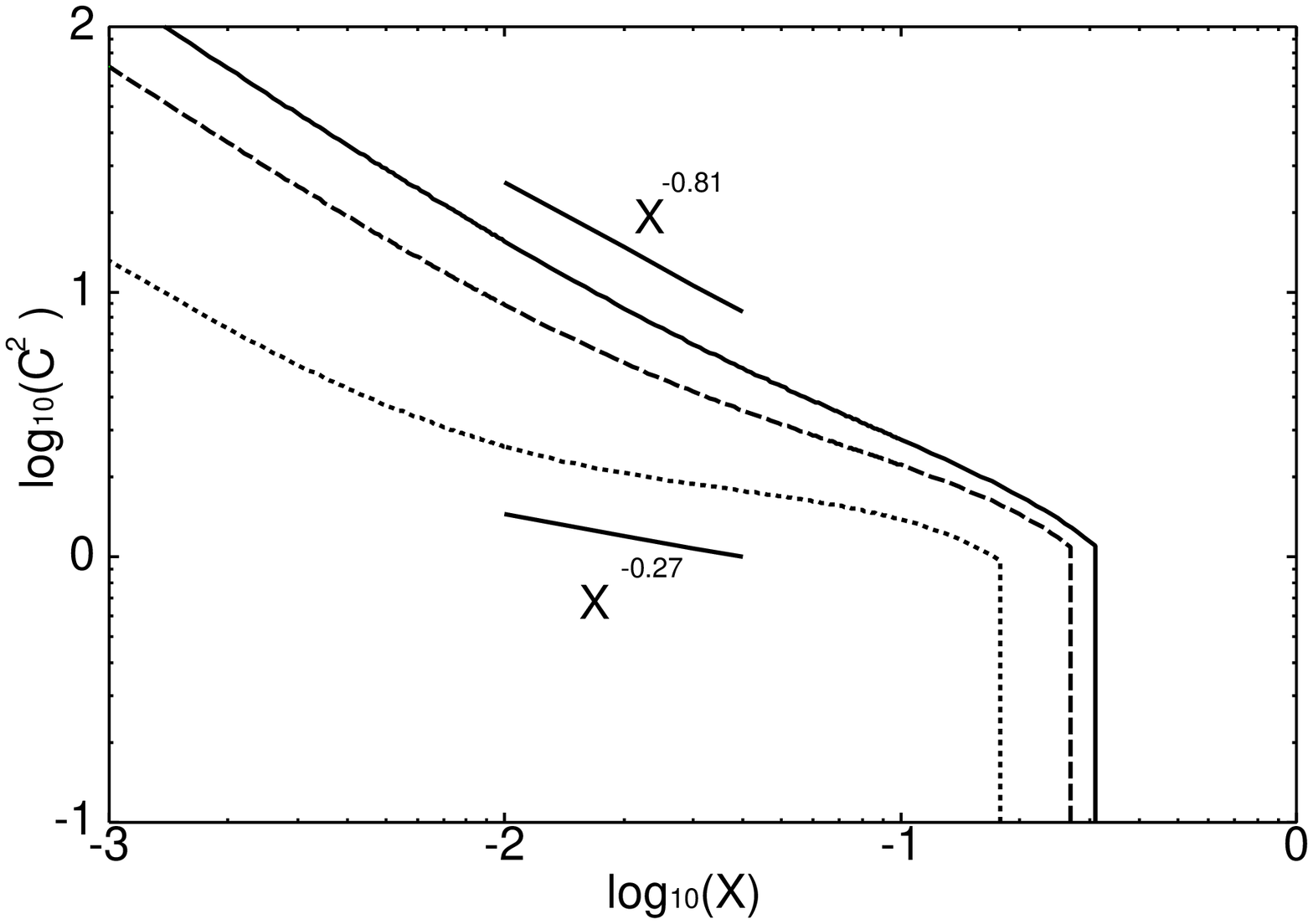}}%
\resizebox{8cm}{!}{\includegraphics{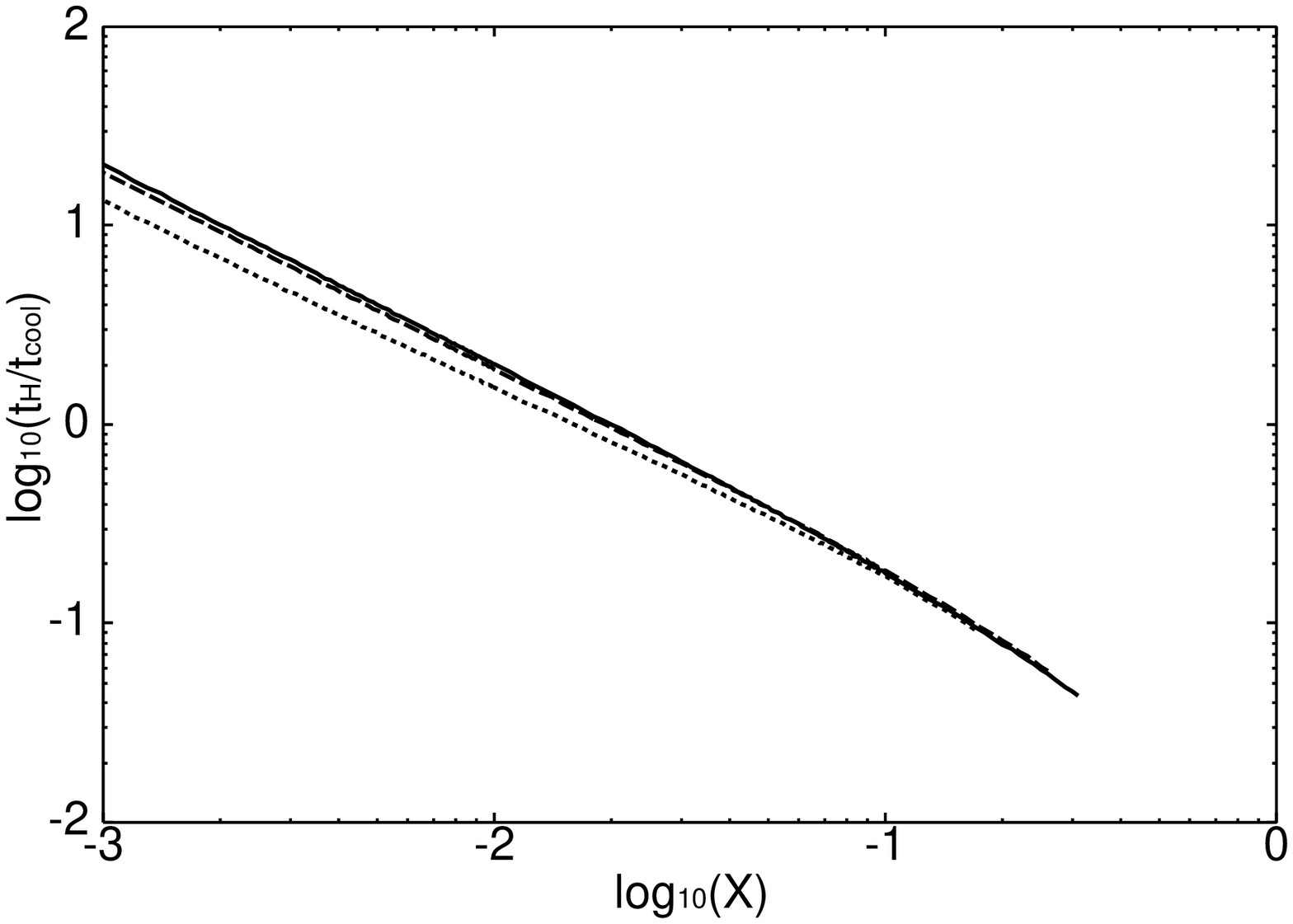}}\\
\resizebox{8cm}{!}{\includegraphics{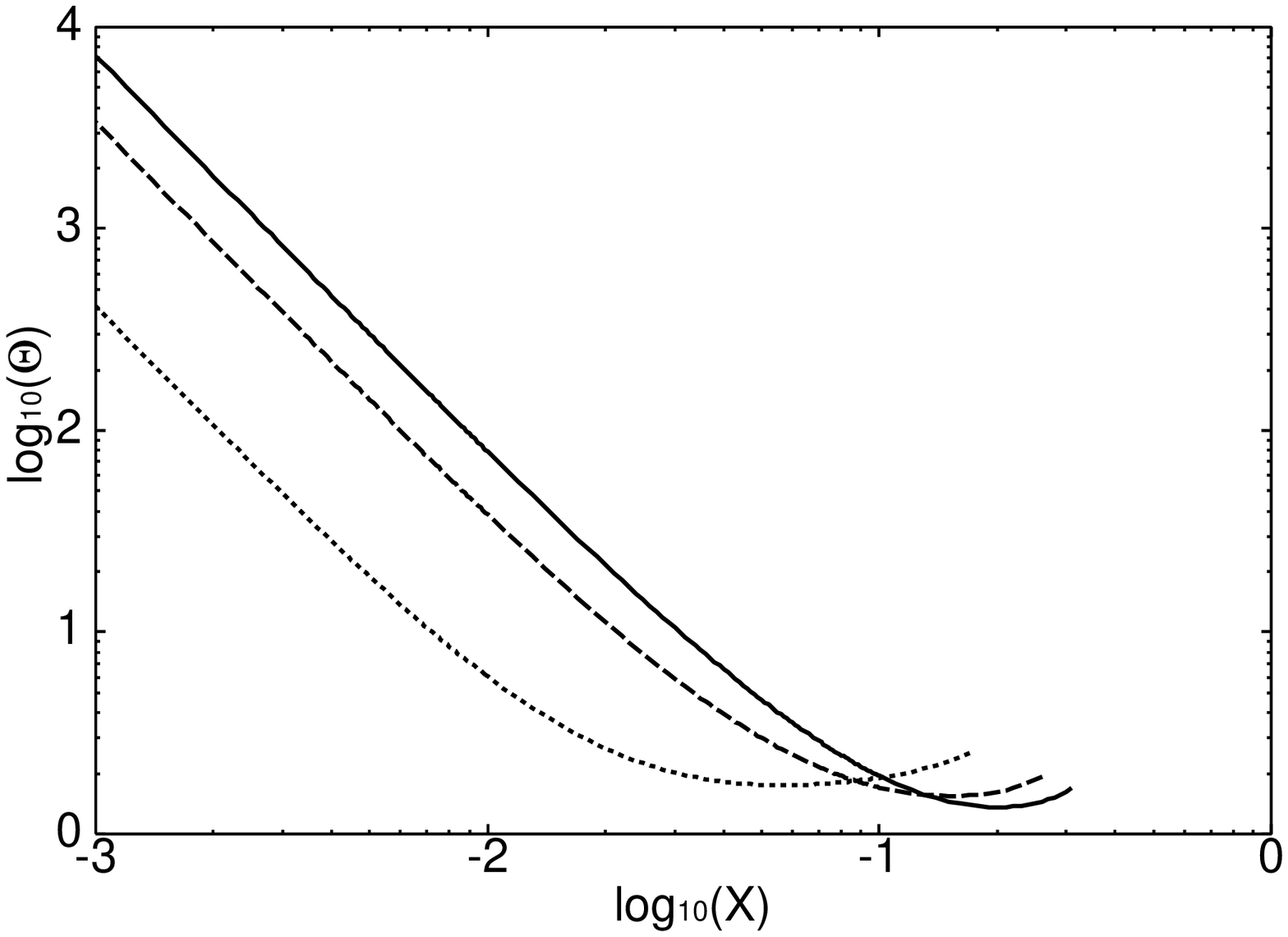}}%
\resizebox{8cm}{!}{\includegraphics{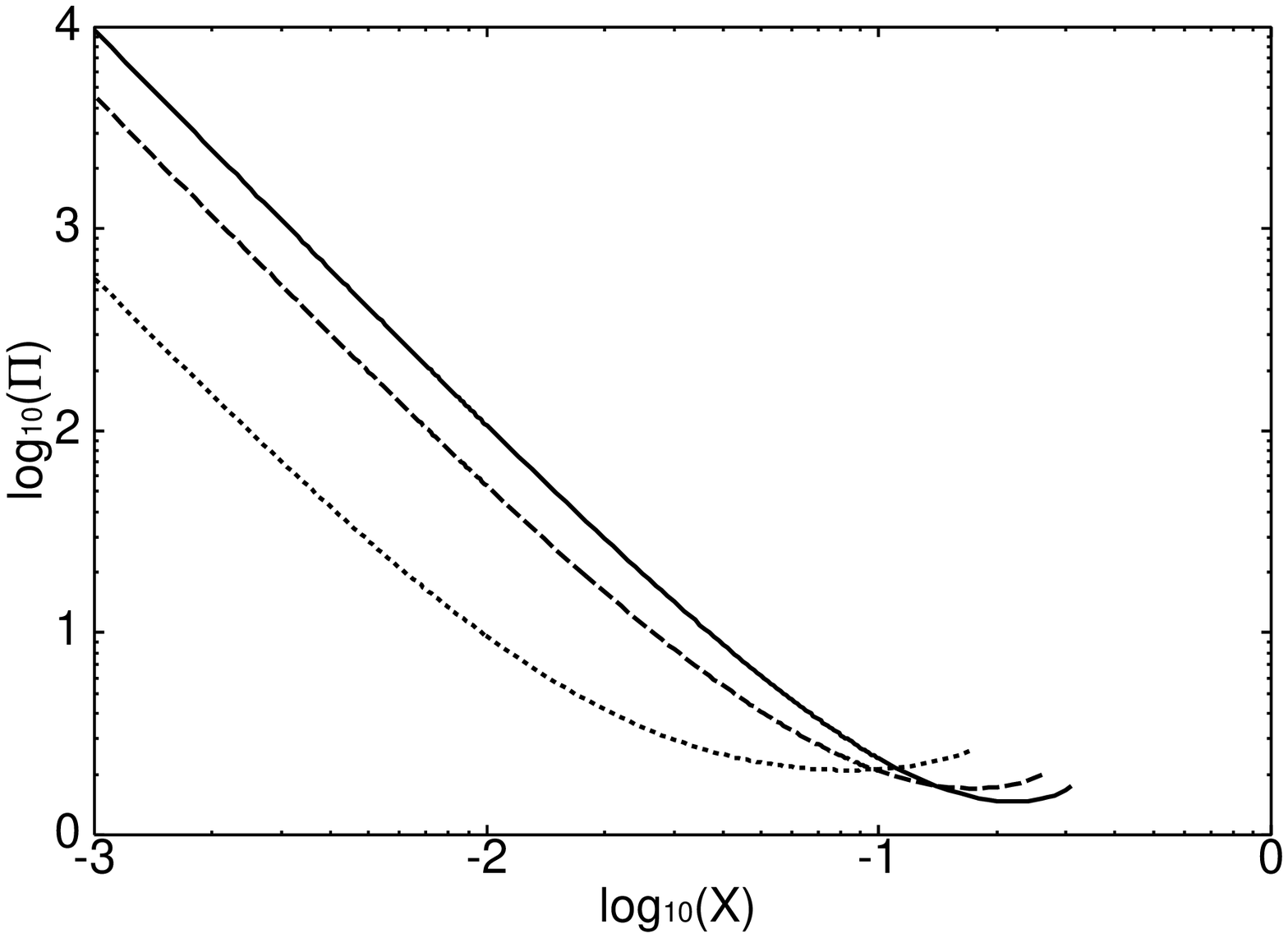}}\\
\ \\
\caption{Similarity variables with $A$=3/2, $B$=1, $K_{0}$=0.01, 
and ${\mathcal{H}}$=0. 
Each line shows the solutions of $\epsilon$=1 (solid line), 
$\epsilon$=2/3 (dashed line), and $\epsilon$=1/3 (dotted line), respectively. 
Each panel indicates radial profiles of the density (top left), 
the proper velocity (top right),
the square of the sound speed corresponding to the temperature (middle left), 
the ratio of the Hubble time to the cooling time (middle right), 
the change of internal energy on the unit mass (bottom left), and 
the work done on the unit mass (bottom right), respectively.}
\label{fig:figure4}
\end{figure*}%
\begin{figure*}
\resizebox{8cm}{!}{\includegraphics{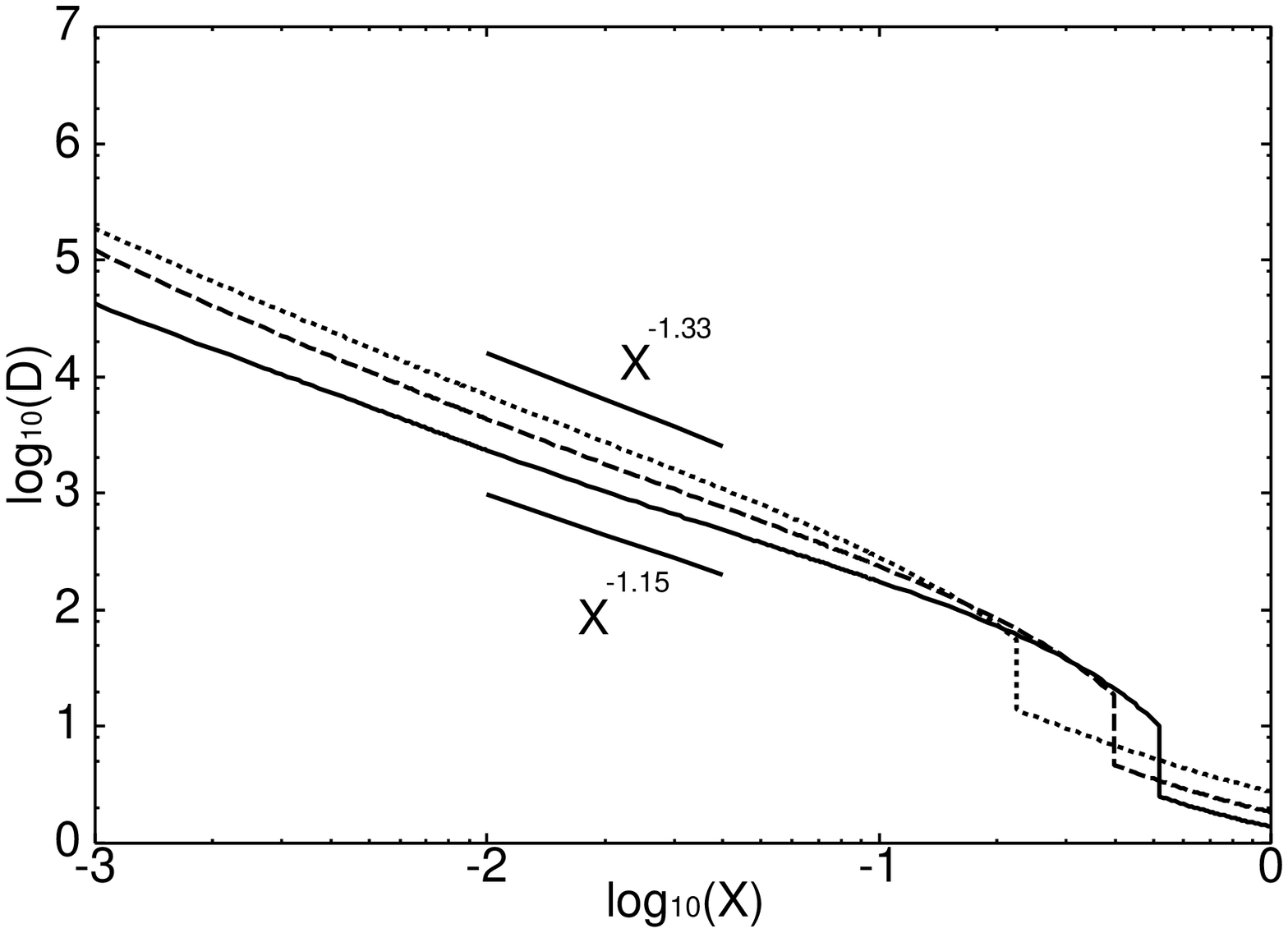}}%
\resizebox{8cm}{!}{\includegraphics{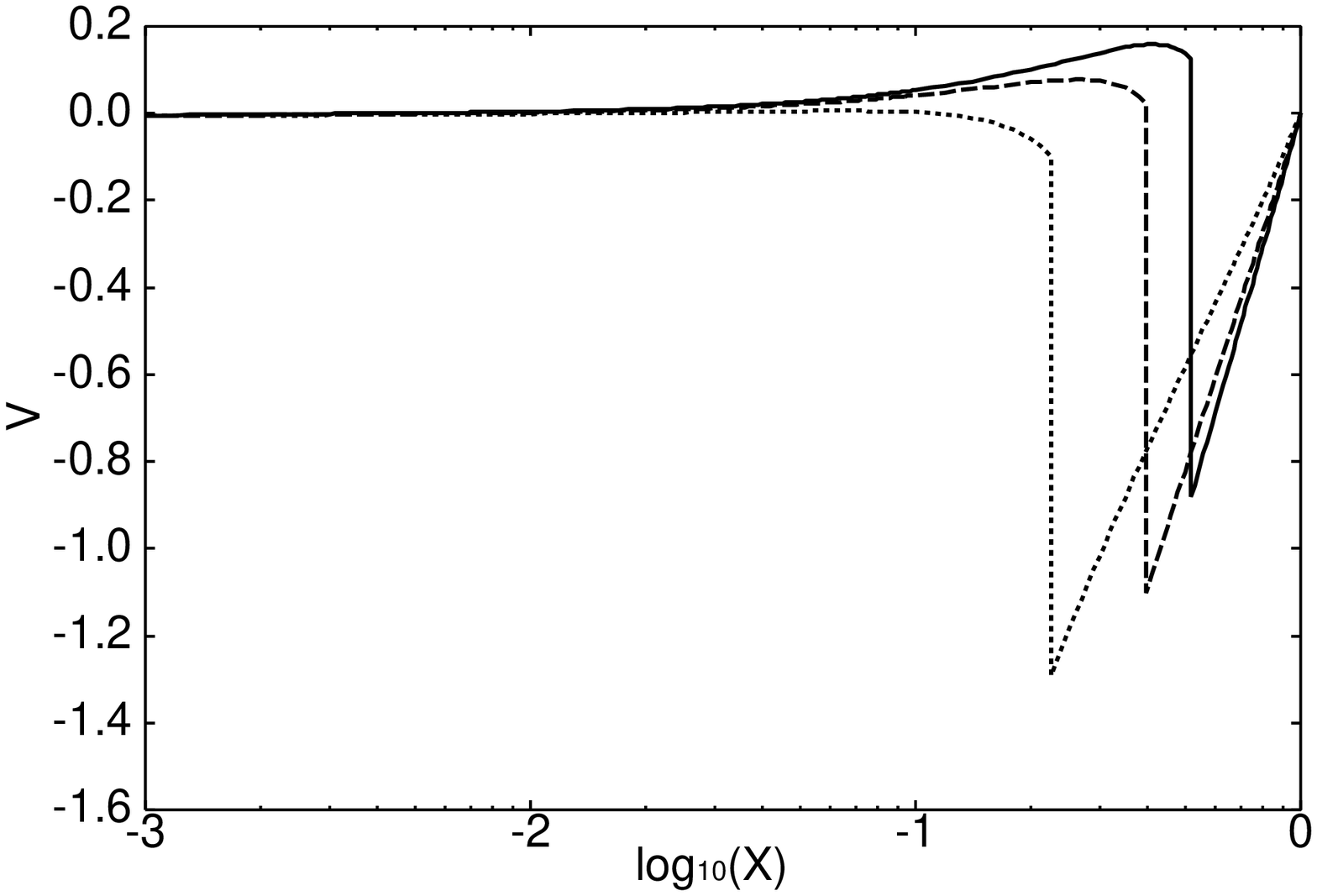}}\\
\resizebox{8cm}{!}{\includegraphics{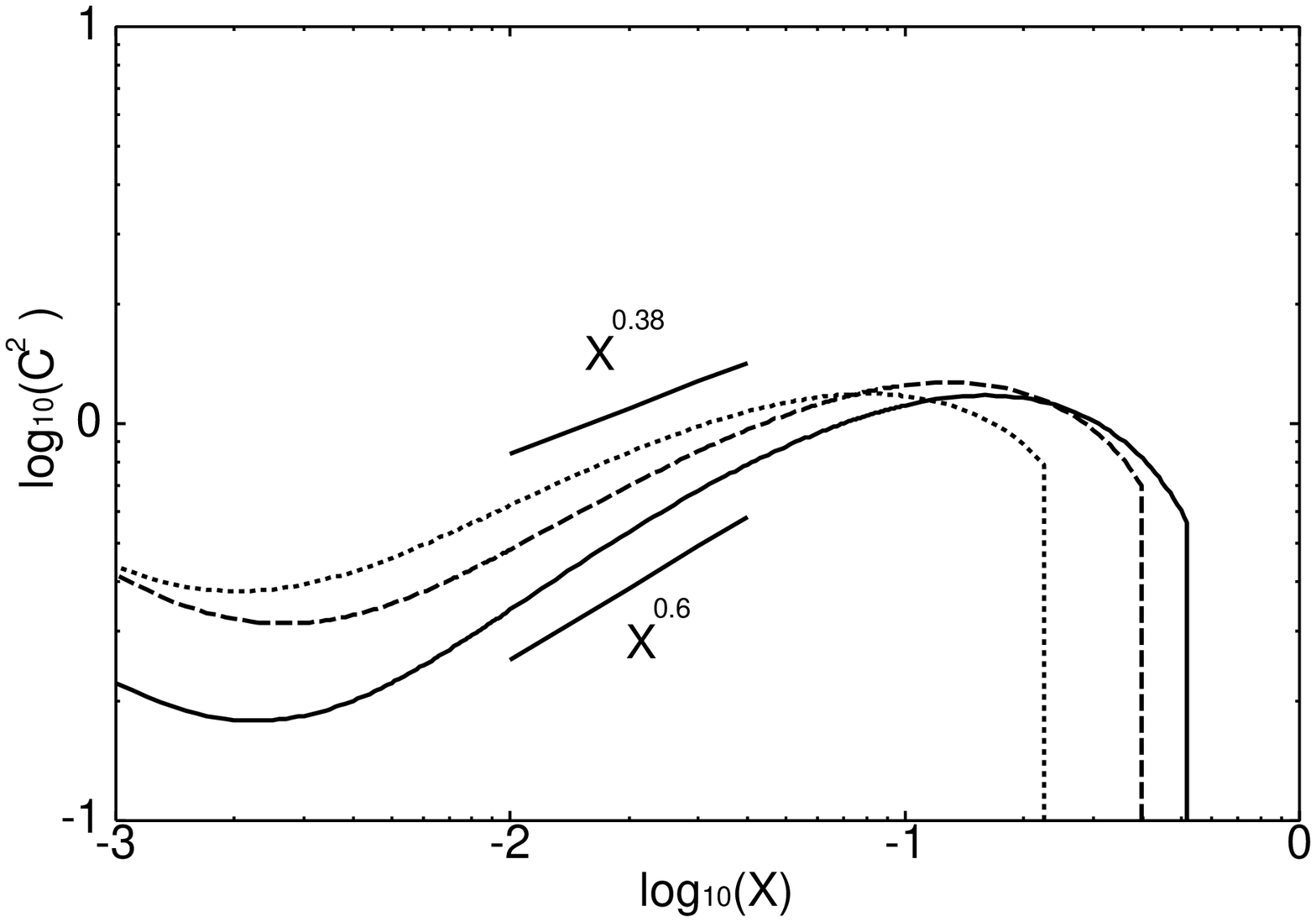}}%
\resizebox{8cm}{!}{\includegraphics{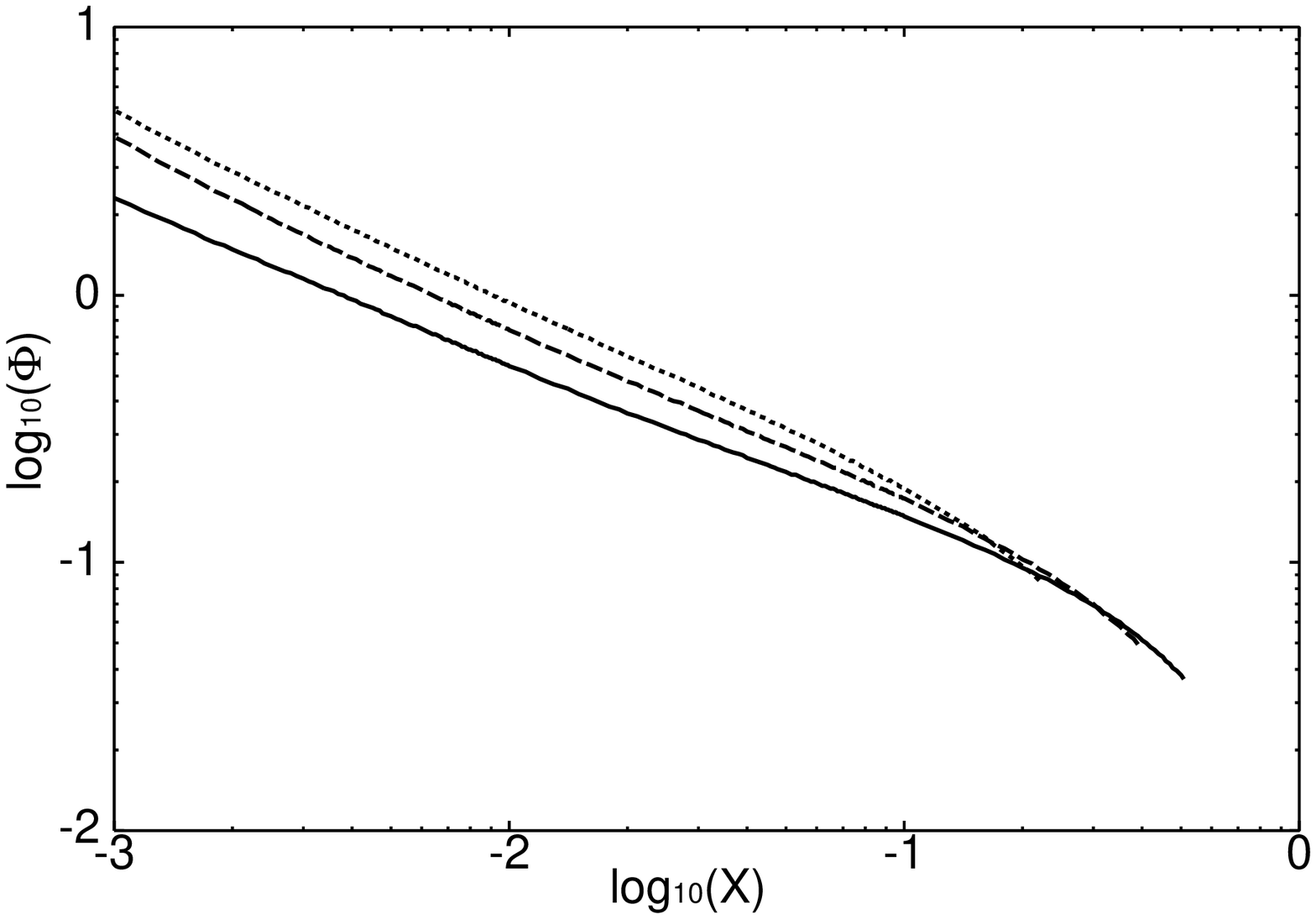}}\\
\resizebox{8cm}{!}{\includegraphics{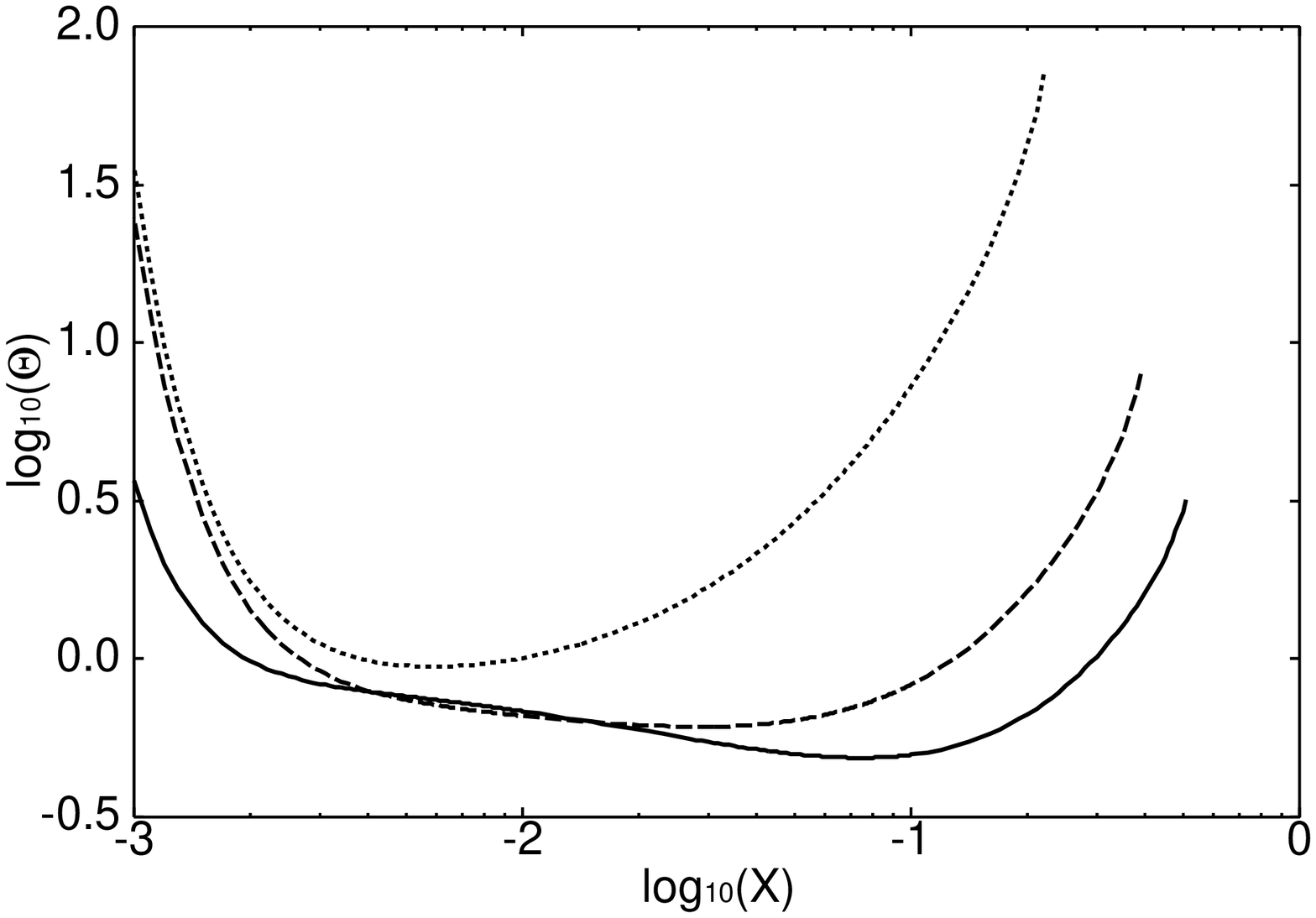}}%
\resizebox{8cm}{!}{\includegraphics{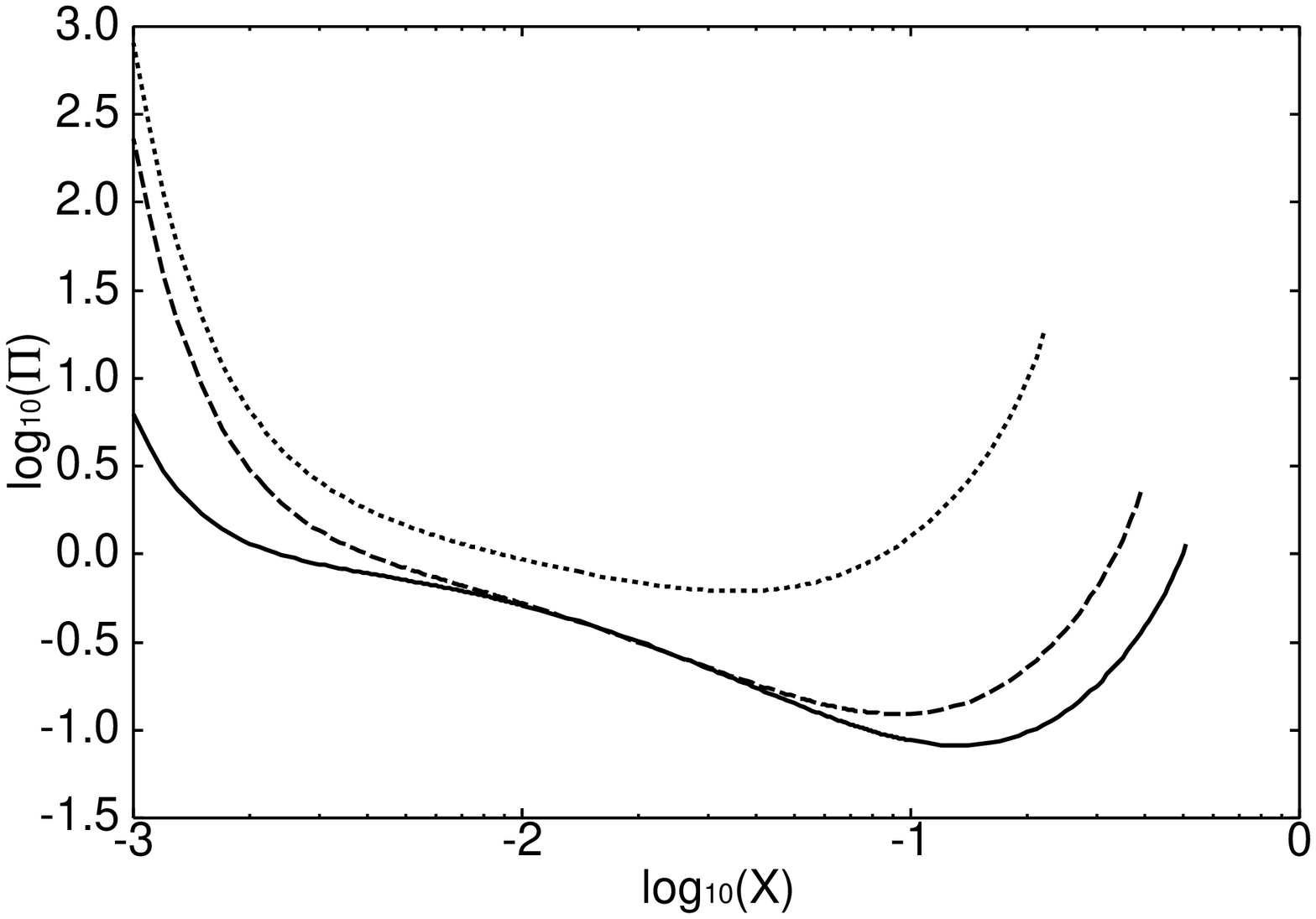}}\\
\ \\
\caption{Similarity variables with $A$=3/2, $B$=1, $K_{0}$=0.01, 
and ${\mathcal{H}}$=0.8. 
The middle right panel indicates the ratio of the cooling rate to the heating 
rate, and the other panels are the same as in Fig. 4.}
\label{fig:figure5}
\end{figure*}%
\begin{table}
  \centering
  \begin{center}
  \begin{tabular}{cccccc}\\
  \hline
   $\epsilon$ & ${\mathcal{H}}$ & $X_{s}$ & $u$ & $v$ & $w$\\
  \hline
  \hline
  \multicolumn{6}{c}{$K_{0}$=0.01}\\
  \hline
     & 0   & 0.311328 & $-$2.03 & $-$0.81 & $-$0.02  \\
  1.0& 0.4 & 0.385248 & $-$1.88 & $-$0.64 & $-$0.36  \\
     & 0.8 & 0.519181 & $-$1.15 & $+$0.60 & $+$1.12 \\  
  \hline
     & 0   & 0.269172 & $-$2.00 & $-$0.67 & $+$0.02\\
  2/3& 0.4 & 0.320795 & $-$1.79 & $-$0.41 & $-$0.42\\
     & 0.8 & 0.397242 & $-$1.26 & $+$0.50 & $+$1.27\\
  \hline
     & 0   & 0.179164 & $-$1.83 & $-$0.27 & $+$0.15 \\
  1/3& 0.4 & 0.199747 & $-$1.64 & $+$0.03 & $-$0.02 \\
     & 0.8 & 0.224469 & $-$1.33 & $+$0.38 & $+$2.06 \\
  \hline
  \hline
  \multicolumn{6}{c}{$K_{0}$=0.1}\\
  \hline
     & 0   & 0.185853 & $-$2.01 & $-$0.97 & $+$0.002 \\
  1.0& 0.4 & 0.209495 & $-$1.99 & $-$0.96 & $+$0.03 \\
     & 0.8 & 0.247105 & $-$1.96 & $-$0.94 & $+$0.06 \\  
  \hline
     & 0   & 0.154968 & $-$2.00 & $-$0.94 & $-$0.001 \\
  2/3& 0.4 & 0.171938 & $-$1.98 & $-$0.93 & $-$0.03 \\
     & 0.8 & 0.197048 & $-$1.95 & $-$0.92 & $-$0.06 \\
  \hline
     & 0   & 0.099695 & $-$1.97 & $-$0.88 & $+$0.003 \\
  1/3& 0.4 & 0.107732 & $-$1.94 & $-$0.87 & $+$0.02 \\
     & 0.8 & 0.118345 & $-$1.92 & $-$0.85 & $-$0.06 \\
  \hline
  \hline
  \end{tabular}
\caption{Properties of solutions for $\Lambda \propto\rho^{3/2}T$.
${\mathcal{H}}$ is the parameter of heating, $w$ is the slope of the velocity 
fitted from $X$=0.01 to 0.04, and the other parameters 
are the same as those given in Table 1.} 
\label{tab:Table2}
\end{center}
\end{table}%
We investigate the properties of solutions for $\Lambda \propto \rho^{3/2}T$, 
which is the same cooling function as that of Abadi et al.(2000). 
Figure 4 shows the resultant profiles for $K_{0}$=0.01 
without heating, ${\mathcal{H}}$=0. 
Each line shows the results of $\epsilon$=1 (solid line), 
$\epsilon$=2/3 (dashed line), and $\epsilon$=1/3 (dotted line), respectively. 
Table 2 gives the shock radius $X_{s}$, and the slopes of the 
similarity variables which were fitted from $X$=0.01 to 0.04. 
Outside of the shock radius, the gas pressure is assumed to be zero. 
Thus, cooling and heating are neglected in this region. 
The fluid variables have the same shapes as the adiabatic similarity 
solution in Fig. 3. For $X<X$$_{s}$, the gas has 
almost hydrostatic distributions. However, in the inner region, 
where the radiative cooling time is less than the age, 
radiative cooling affects the gas distribution. As a result, 
the pressure support decreases, and a cooling inflow is established.  
In Fig. 4, the non-dimensional changes of the internal energy and 
the work done on a unit mass are shown, which are defined by 
\begin{eqnarray}
\frac{dE}{dt}=\frac{r_{ta}^{2}}{t^{3}_{H}} \Theta,
\label{eq:equation45-1}
\end{eqnarray}
\begin{eqnarray}
\frac{-Pd(1/\rho)}{dt}=\frac{r_{ta}^{2}}{t^{3}_{H}} \Pi,
\label{eq:equation45-2}
\end{eqnarray}
where $P$ is the gas pressure. 
Using equations (\ref{eq:equation18})-(\ref{eq:equation21}), 
the non-dimensional variables are given by 
\begin{eqnarray}
\Theta=\frac{2C^{2}}{\gamma(\gamma-1)}[\xi-1+\frac{V-\xi X}{C}\frac{dC}{dX}],
\label{eq:equation47}
\end{eqnarray}
\begin{eqnarray}
\Pi=\frac{C^{2}}{\gamma}[\frac{V-\xi X}{D}\frac{dD}{dX}-2].
\label{eq:equation46}
\end{eqnarray}
In Fig. 4, the gas is compressed because $\Pi>$0. More than 60 percent of 
the work is turned into internal energy, and 
the gas heats up. Since $\Theta$ increases with decreasing radius, 
the temperature increases with decreasing radius. 

We discuss the slopes of the similarity variables, including cooling. 
In Fig. 4, the density slopes of $\epsilon$=1 and 2/3 are shallower than 
those of Fig. 3. For $\epsilon$=1/3, the density is steeper than the result 
of Fig. 3. On the other hand, the temperature increases with decreasing 
radius in all cases. 
Abadi et al.(2000) found that the slopes of the density and temperature 
approach $-$2 and $-$1 as $X$$\rightarrow$0, respectively.  
In the case of ($A$,$B$), assuming that the infall velocity 
is $V$$<<$$-\xi X$ and approach $V$=$V_{0}X^{w}$ as $X$$\rightarrow$0, 
the density and sound speed 
are expressed from equations (\ref{eq:equation27})-(\ref{eq:equation29}), and 
(\ref{eq:equation29-2}) as follows:
\begin{equation}
D=D_{0}X^{s},\nonumber\\
C=C_{0}X^{t},\nonumber\\
\label{eq:equation46-1}
\end{equation}
where 
\begin{eqnarray}
s&=&\frac{w+B-2}{A-1} \hspace{0.4cm} (A\neq 1),\nonumber\\
 &=&B-4 \hspace{1.1cm} (A=1),
\label{eq:equation46-2}
\end{eqnarray}
\begin{eqnarray}
t&=&-\frac{1}{2}.
\label{eq:equation46-3}
\end{eqnarray}
From Table 2, in the case of $K_{0}$=0.1, 
the infall velocity is approximately constant (i.e.,$w$=0). 
Substituting $A$=3/2, $B$=1, and $w$=0 for equation 
(\ref{eq:equation46-2}), 
the density slope becomes $-$2, and the temperature slope is $-1$ 
from equation (\ref{eq:equation46-3}). We can see form Table 2 that 
the density and temperature slopes of numerical calculations 
for $K_{0}$=0.1 are considerably close to $-$2 and $-1$, respectively. 
In the case of $K_{0}$=0.01 without heating, 
the infall velocity is approximately constant. 
However, $V$$<<$$-\xi X$ is not fully satisfied because the energy loss 
is smaller than that of $K_{0}$=0.1, in which case the gas distribution 
approaches the adiabatic one. Thus, the density and temperature slopes turn 
into the middle gradient between the case of $K_{0}$=0.1 and the 
adiabatic solution in Fig. 3. 

Next, we represent the similarity solutions that incorporate the effects of 
heating and cooling. 
Figure 5 shows the resultant profiles for ${\mathcal{H}}$=0.8, 
and the other parameters have the same values as in the previous model. 
We show the ratio $\Phi$ of the cooling rate to the heating rate, 
\begin{eqnarray}
\Phi=\frac{\Lambda}{\Gamma}
=\frac{K_{0}}{{\mathcal{H}}\gamma(\gamma-1)}D^{A-1}C^{2B-2}.
\label{eq:equation48}
\end{eqnarray}
Since the heating rate is greater than the cooling rate behind the shock,
the infall velocity is weakened by heating, and the gas expands because 
$\Pi<$0 and cools down as $\Theta<$0. 
The slope of the density becomes substantially 
shallow compared with the result of no heating, shown in Fig. 4,  
because the concentration of mass is suppressed due to heating. 
On the other hand, temperature increases with decreasing radius for 
$X$$>$0.1. However, the change of internal energy ($\Theta$) steeply 
decreases with decreasing radius, except in the central region. 
As a result, in $X$=0.003 $\sim$ 0.1, the temperature becomes 
decreasing profiles toward the center. 
When the flow approaches the central region, 
the cooling rate becomes greater than the heating rate. 
In this region, radiative cooling produces a cooling inflow, 
the velocity becomes negative and the gas is compressed. 
Because the change of the internal energy ($\Theta$) shows a steeper rise 
toward the center in $X$$<$0.003, the gas heats up and the temperature shows 
increasing profiles toward the origin. 

For the case of $K_{0}$=0.1, 
we can see from Table 2 that the slopes of the density and the 
temperature do not change significantly by increasing the coefficient 
of heating, ${\mathcal{H}}$, and are about $-$2 and $-$1, 
respectively. The reason is that the flows are not affected by heating 
because the cooling is stronger than heating at all radii.

\subsection{Cases of $B$=1/2, or $-$1/2}
\begin{figure}
\resizebox{8cm}{!}{\includegraphics{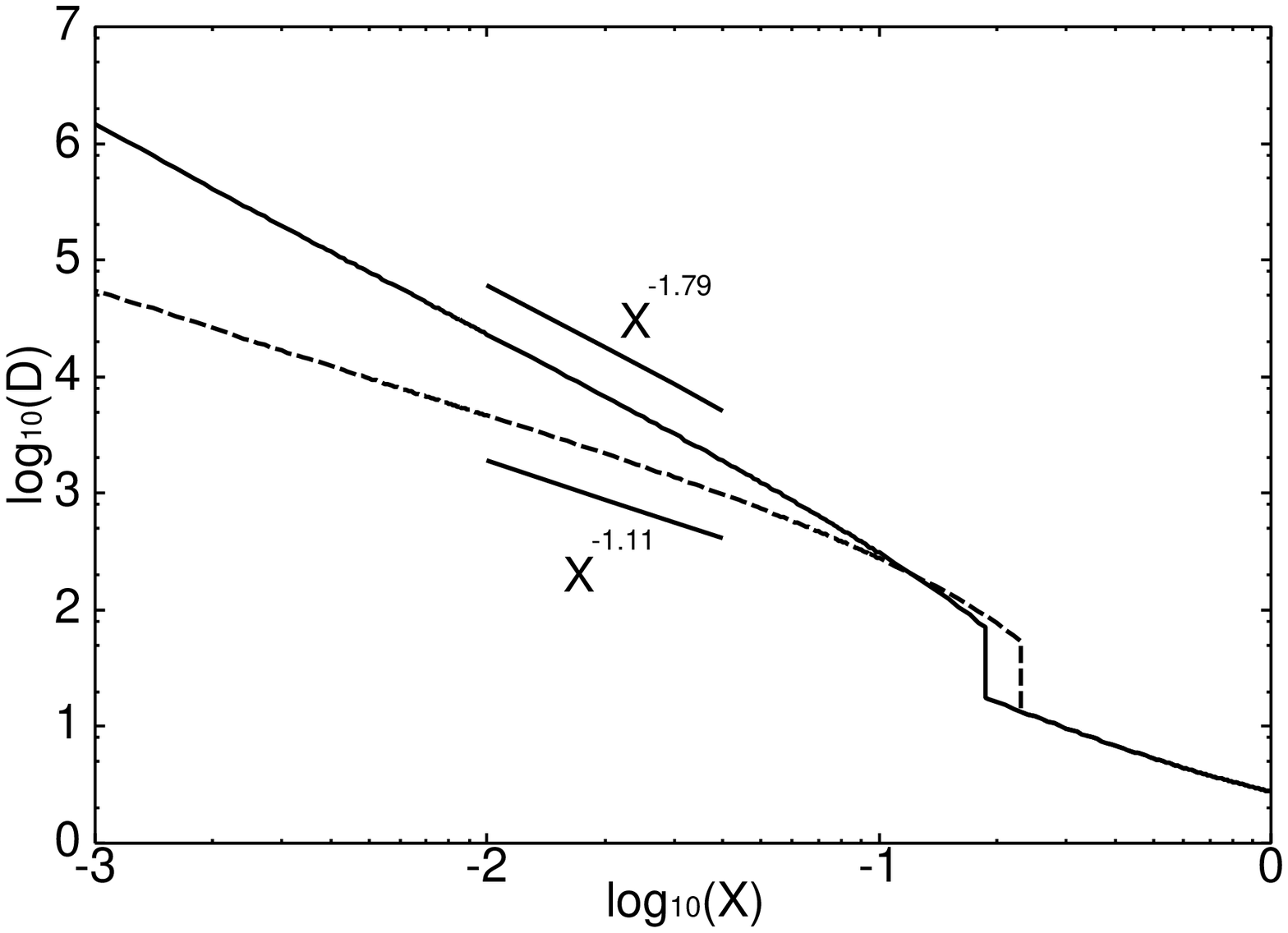}}\\
\resizebox{8cm}{!}{\includegraphics{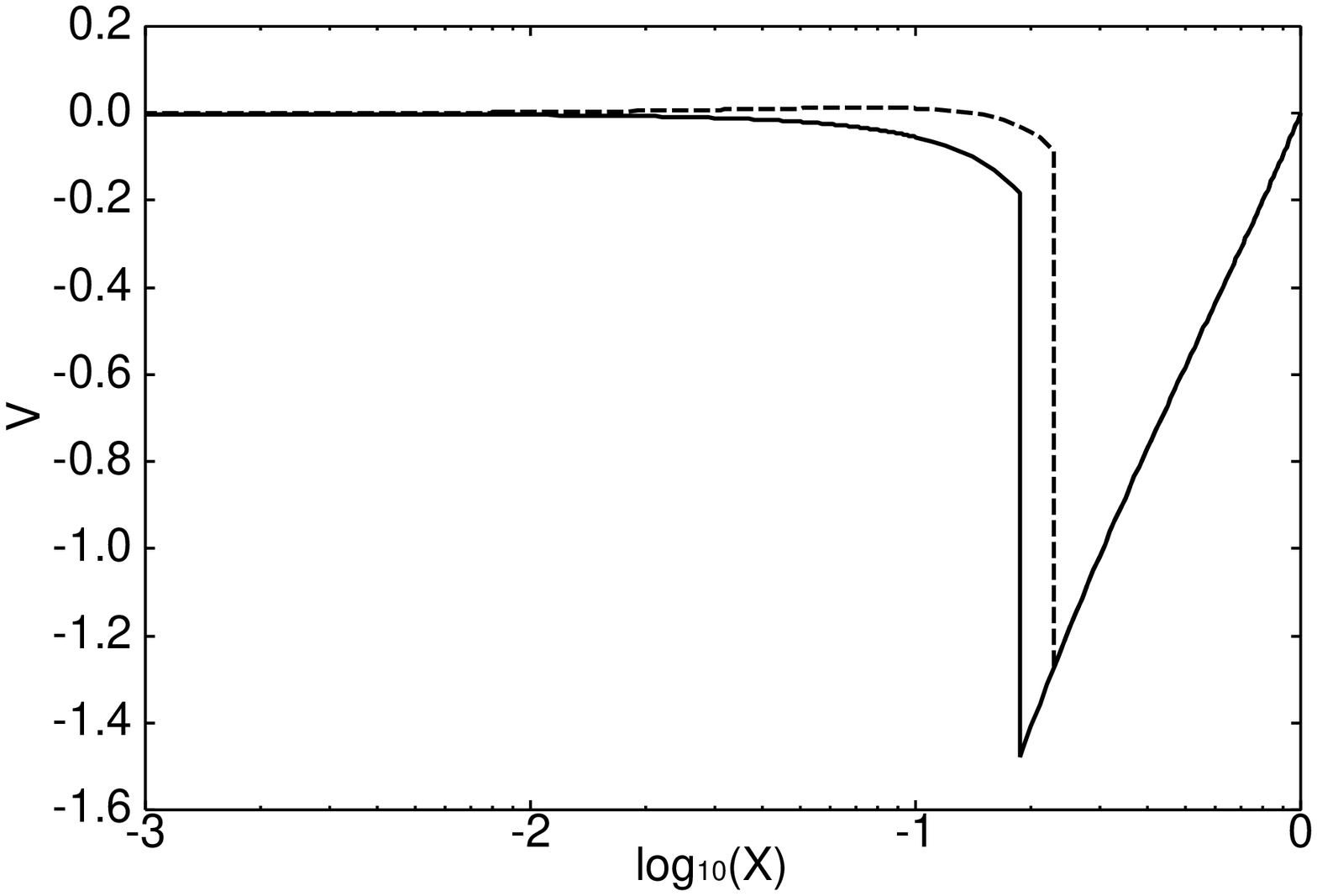}}\\
\resizebox{8cm}{!}{\includegraphics{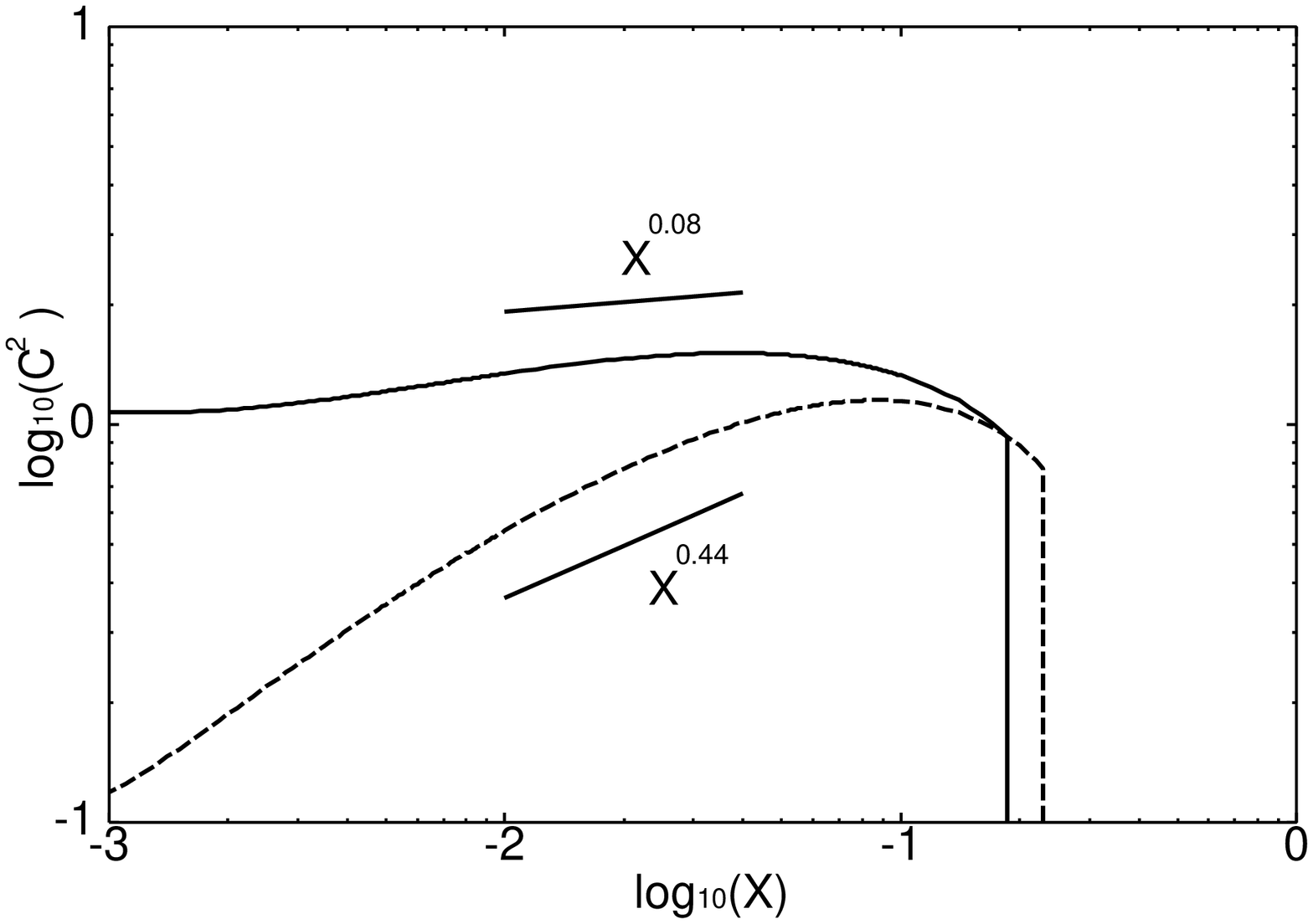}}%
\ \\
\caption{Similarity variables with $\epsilon$=1/3($n$=$-$1), 
$A$=4/3, $B$=1/2, and 
$K_{0}$=0.01. The solid and dashed lines indicate the results 
without heating, ${\mathcal{H}}$=0, and with heating, ${\mathcal{H}}$=0.8, 
respectively. Each panel represents radial profiles of the density 
(top panel), the proper velocity (middle panel), and 
the square of the sound speed corresponding to the temperature 
(bottom panel), respectively.}
\label{fig:figure6}
\end{figure}%
\begin{figure}
\resizebox{8cm}{!}{\includegraphics{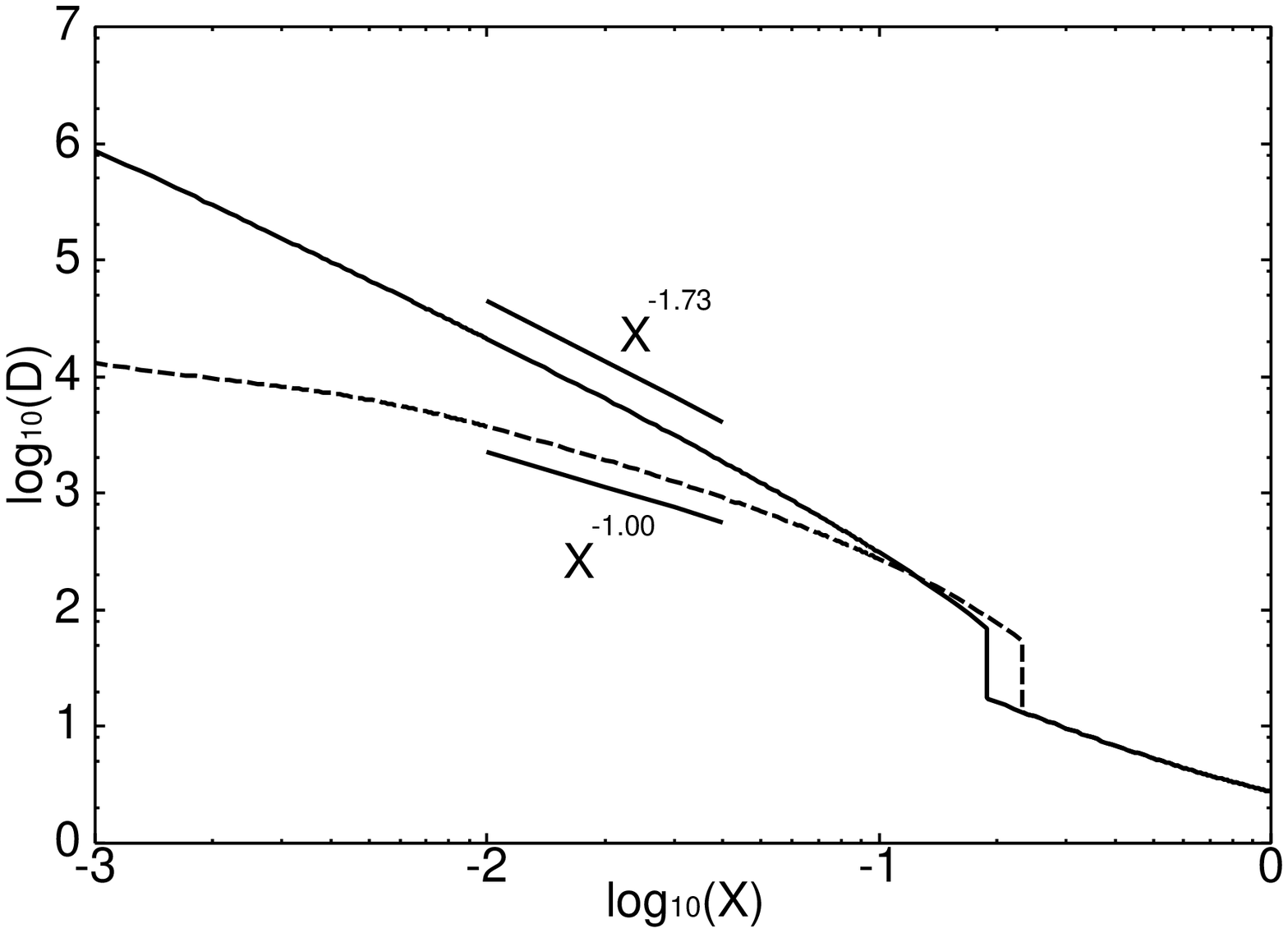}}\\
\resizebox{8cm}{!}{\includegraphics{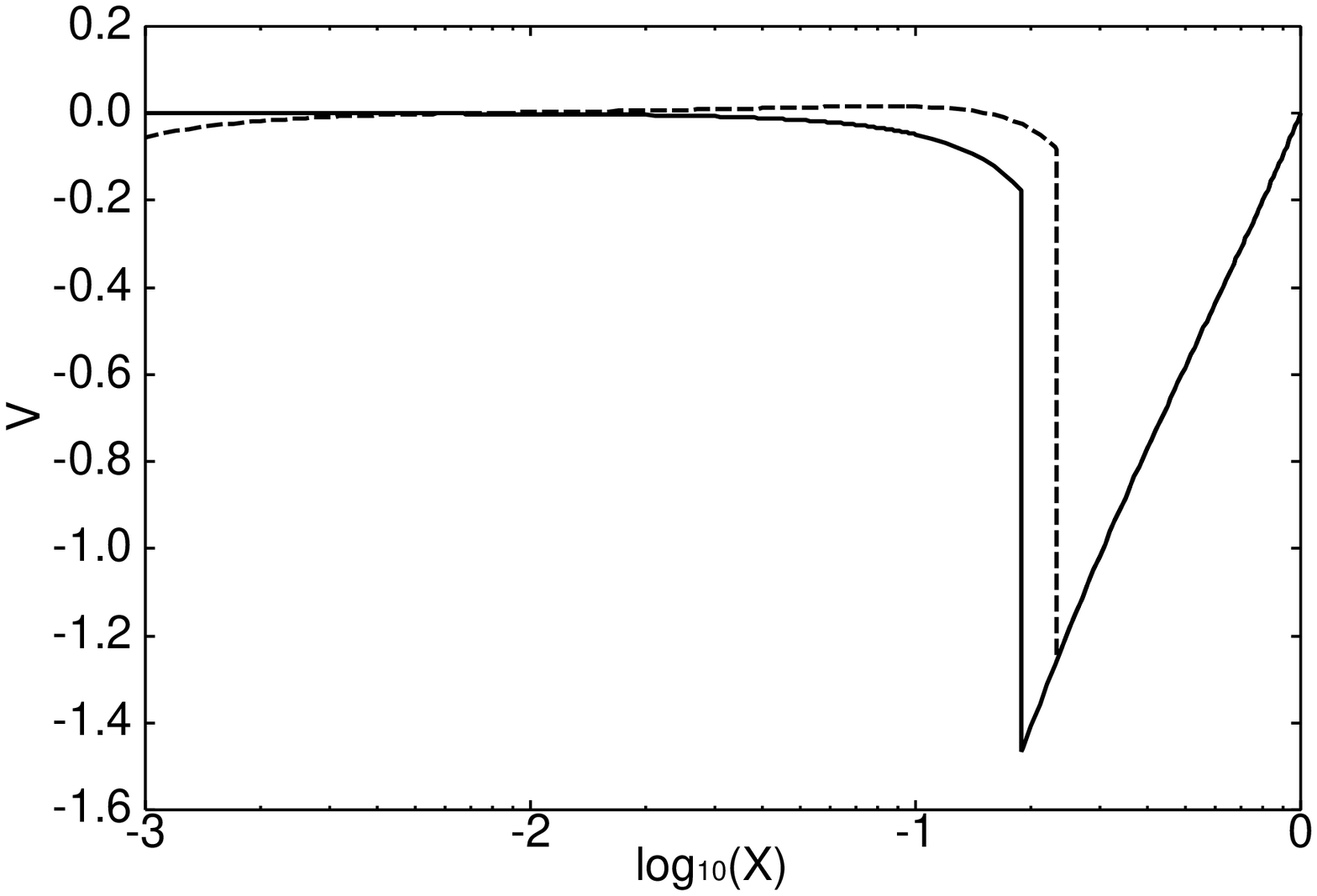}}\\
\resizebox{8cm}{!}{\includegraphics{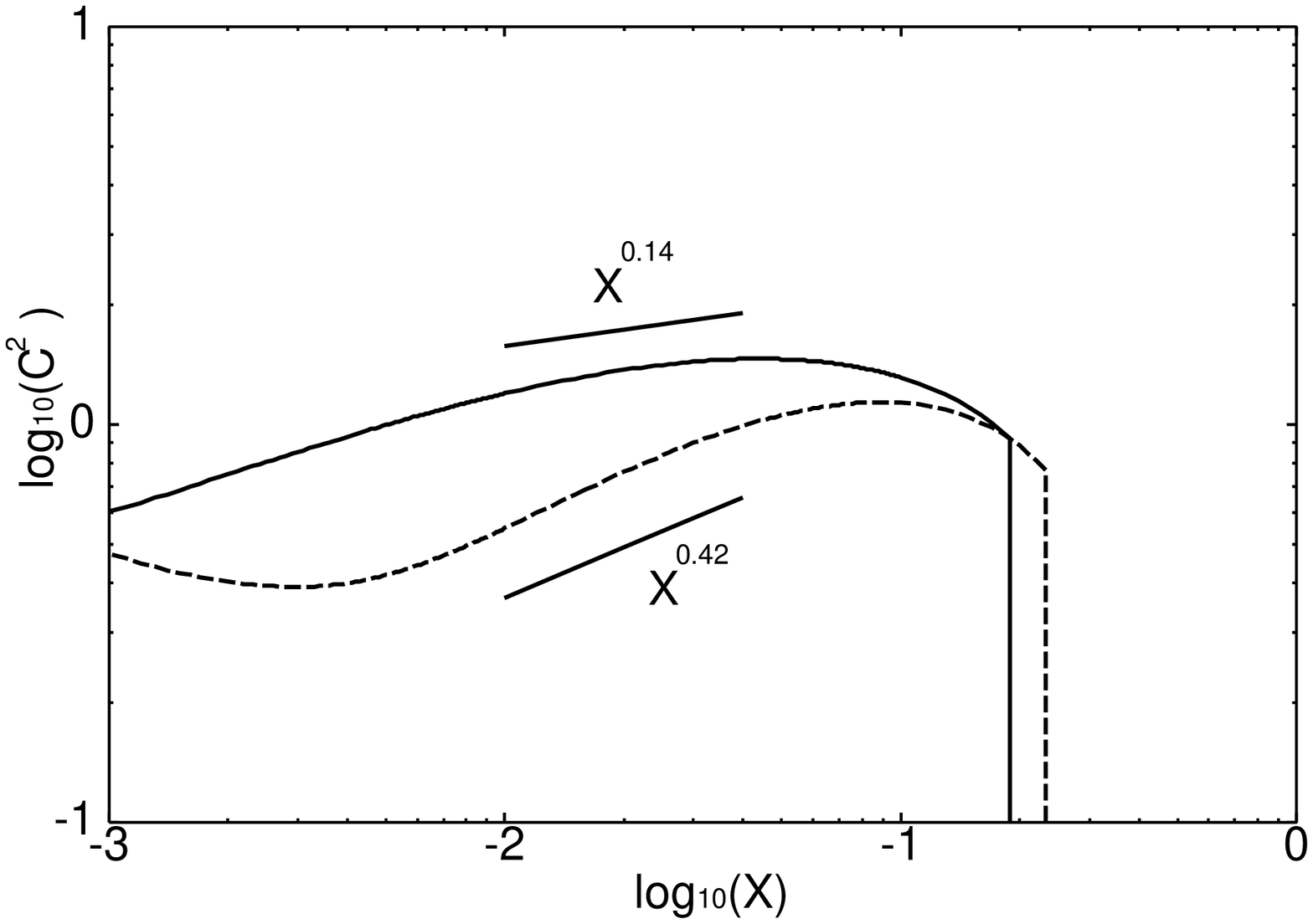}}%
\ \\
\caption{Same as in Fig. 6, except $A$=1 and $B$=$-$1/2.}
\label{fig:figure7}
\end{figure}%
\begin{table}
\centering
\begin{center}
\begin{tabular}{ccccccc}\\
\hline
   $\epsilon$ & $A$,$B$ &${\mathcal{H}}$ & $X_{s}$ & $u$ & $v$ & $w$ \\
\hline
\hline  
   \multicolumn{7}{c}{$K_{0}$=0.01}\\
\hline
\hline  
  1.0&14/9,1/2 &0   & 0.318480 & $-$2.19 & $-$0.69 & $+$0.16 \\
     &         &0.8 & 0.512580 & $-$1.34 & $+$0.34 & $+$2.52 \\  
\hline
  2/3& 3/2,1/2 &0   & 0.277630 & $-$2.11 & $-$0.44 & $+$0.24 \\
     &         &0.8 & 0.397755 & $-$1.26 & $+$0.46 & $+$1.64 \\
\hline
  1/3& 4/3,1/2 &0   & 0.186403 & $-$1.79 & $+$0.08 & $+$1.06  \\
     &         &0.8 & 0.229463 & $-$1.11 & $+$0.44 & $+$0.86 \\
\hline
\hline
  1.0& 5/3,-1/2&0   & 0.323424 & $-$2.31 & $-$0.54 & $+$0.38 \\
     &         &0.8 & 0.496057 & $-$1.71 & $-$0.07 & $-$1.36 \\  
\hline
  2/3& 3/2,-1/2&0   & 0.283470 & $-$2.12 & $-$0.23 & $+$0.55  \\
     &         &0.8 & 0.398342 & $-$1.25 & $+$0.40 & $+$2.61  \\
\hline
  1/3& 1.0,-1/2&0   & 0.188644 & $-$1.73 & $+$0.14 & $+$1.47  \\
     &         &0.8 & 0.232110 & $-$1.00 & $+$0.42 & $+$0.94 \\
\hline
\hline  
\multicolumn{7}{c}{$K_{0}$=0.1}\\
\hline
\hline  
  1.0&14/9,1/2 &0   & 0.230693 & $-$2.23 & $-$0.91 & $+$0.22 \\
     &         &0.8 & 0.307894 & $-$2.14 & $-$0.86 & $+$0.10 \\  
\hline
  2/3& 3/2,1/2 &0   & 0.210950 & $-$2.25 & $-$0.82 & $+$0.29 \\
     &         &0.8 & 0.273013 & $-$2.14 & $-$0.74 & $+$0.13 \\
\hline
  1/3& 4/3,1/2 &0   & 0.166035 & $-$2.10 & $-$0.30 & $+$0.56 \\
     &         &0.8 & 0.201210 & $-$1.76 & $-$0.02 & $+$0.13 \\
\hline
\hline
  1.0& 5/3,-1/2&0   & 0.262117 & $-$2.48 & $-$0.77 & $+$0.54 \\
     &         &0.8 & 0.334214 & $-$2.35 & $-$0.71 & $+$0.32 \\  
\hline
  2/3& 3/2,-1/2&0   & 0.247433 & $-$2.38 & $-$0.52 & $+$0.65 \\
     &         &0.8 & 0.313539 & $-$2.16 & $-$0.36 & $+$0.15 \\
\hline
  1/3& 1.0,-1/2&0   & 0.185988 & $-$1.76 & $+$0.11 & $+$1.36 \\
     &         &0.8 & 0.226802 & $-$1.11 & $+$0.42 & $+$0.88 \\
\hline
\hline
\end{tabular}
\caption{Properties of solutions which fixed the temperature dependence of 
the radiative cooling functions to $B$=1/2 or $-$1/2. 
The other parameters are the same as those given in Table 2.}
\label{tab:Table3}
\end{center}
\end{table}%
We derive the similarity solutions fixed to $B$=1/2 or $-$1/2. 
The former is similar to the temperature 
dependence of free-free emission. Since $A$ is determined by equation
(\ref{eq:equation12}), 
$A$=14/9 for $\epsilon$=1, $A$=3/2 for $\epsilon$=2/3, and $A$=4/3 for 
$\epsilon$=1/3, respectively. 
The latter is similar to the temperature dependence of the line emission. 
Similarly, $A$=5/3 for $\epsilon$=1, 
$A$=3/2 for $\epsilon$=2/3, and $A$=1 for $\epsilon$=1/3, respectively. 
Because the realistic free-free and line emissions are proportional to 
the square of the density, $A$=2, 
the energy loss of the cooling functions assumed here is less than that of 
the realistic one. Figures 6 and 7 show the solutions of ($A$,$B$)=(4/3,1/2) 
and (1,$-$1/2) with  $\epsilon$=1/3($n$=$-$1) and $K_{0}$=0.01, 
respectively. The solid and dashed lines display the solutions without 
heating, ${\mathcal{H}}$=0, and with heating, ${\mathcal{H}}$=0.8, 
respectively. The shock radius $X_{s}$, and the slopes of the similarity 
variables are given in Table 3. For $\epsilon$=1/3, 
the energy loss of cooling is small because 
$A$=3/4 and 1. Thus, the density and temperature without heating are 
almost the same profiles as the adiabatic solutions in Fig. 3. 
For the heating solution with ${\mathcal{H}}$=0.8, 
the slope of the density becomes shallow and the temperature is lower than 
that of no heating, ${\mathcal{H}}$=0. 
The main reason is the same as the previous discussion in Section 3.2. 

From Table 3, for $\epsilon$=1 and 2/3 in the case of $K_{0}$=0.1, 
the density slopes are steeper than those of $\Lambda$$\propto$$\rho^{3/2}T$. 
If the infall velocity is $V$$<<$$-\xi X$, the slope of temperature 
approaches $-$1, from equation (\ref{eq:equation46-3}). 
On the other hand, the density slope depends on the values of $A$, $B$, 
and the slope of velocity $w$ from equation(\ref{eq:equation46-2}).  
For example, substituting ($A$,$B$,$w$)=(14/9,1/2,0.22) and (5/3,$-$1/2,0.54), 
which is the case of $\epsilon$=1 with $K_{0}$=0.1, 
for equation (\ref{eq:equation46-2}), 
the density slopes become $-$2.3 and $-$2.94, respectively. 
The resultant slope of the numerical calculation is shallower 
than this slope because $V$$<<$$-\xi X$ is not fully satisfied.

\subsection{Two-particle processes, A=2}
\begin{figure}
\resizebox{8cm}{!}{\includegraphics{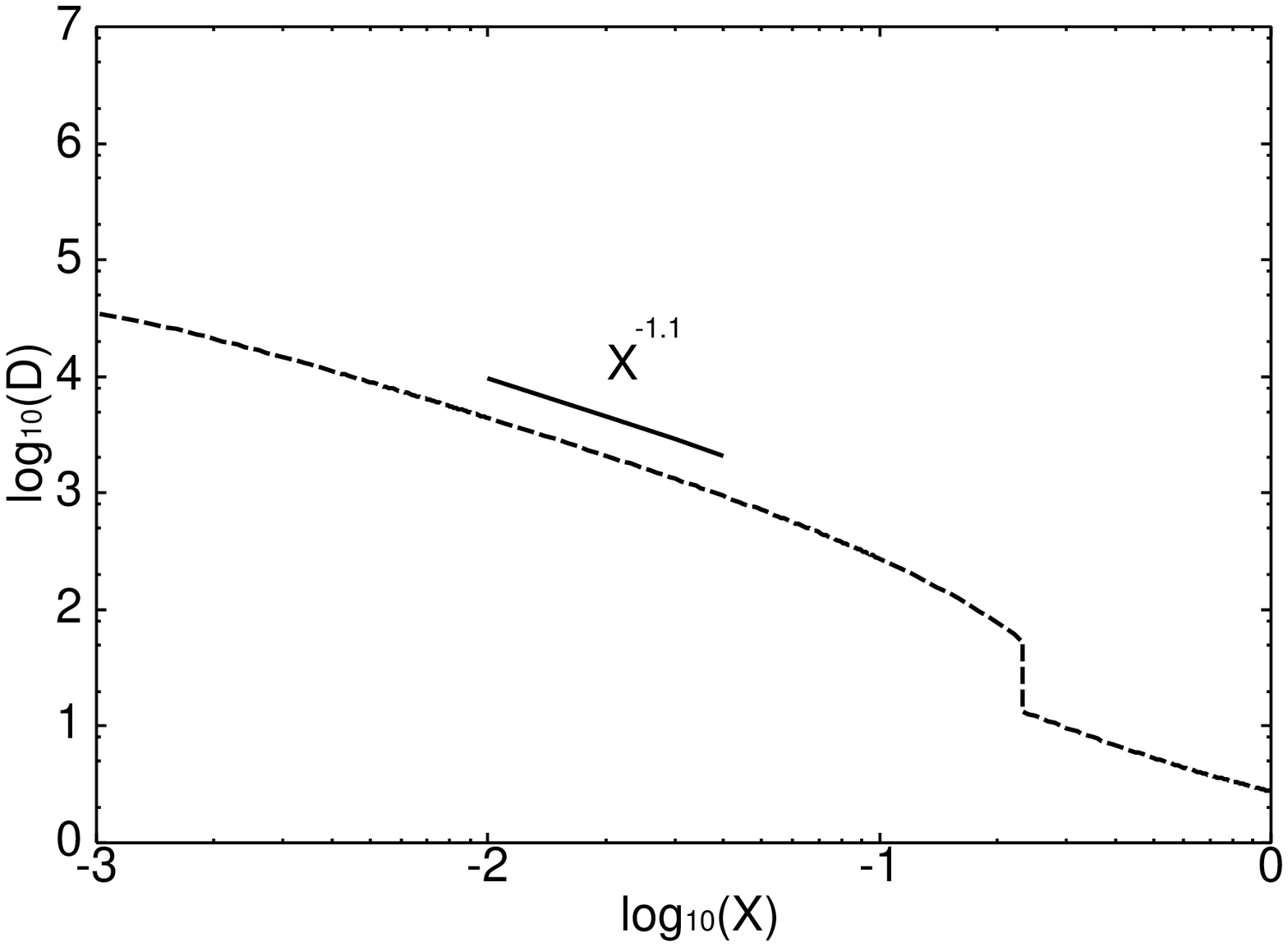}}\\
\resizebox{8cm}{!}{\includegraphics{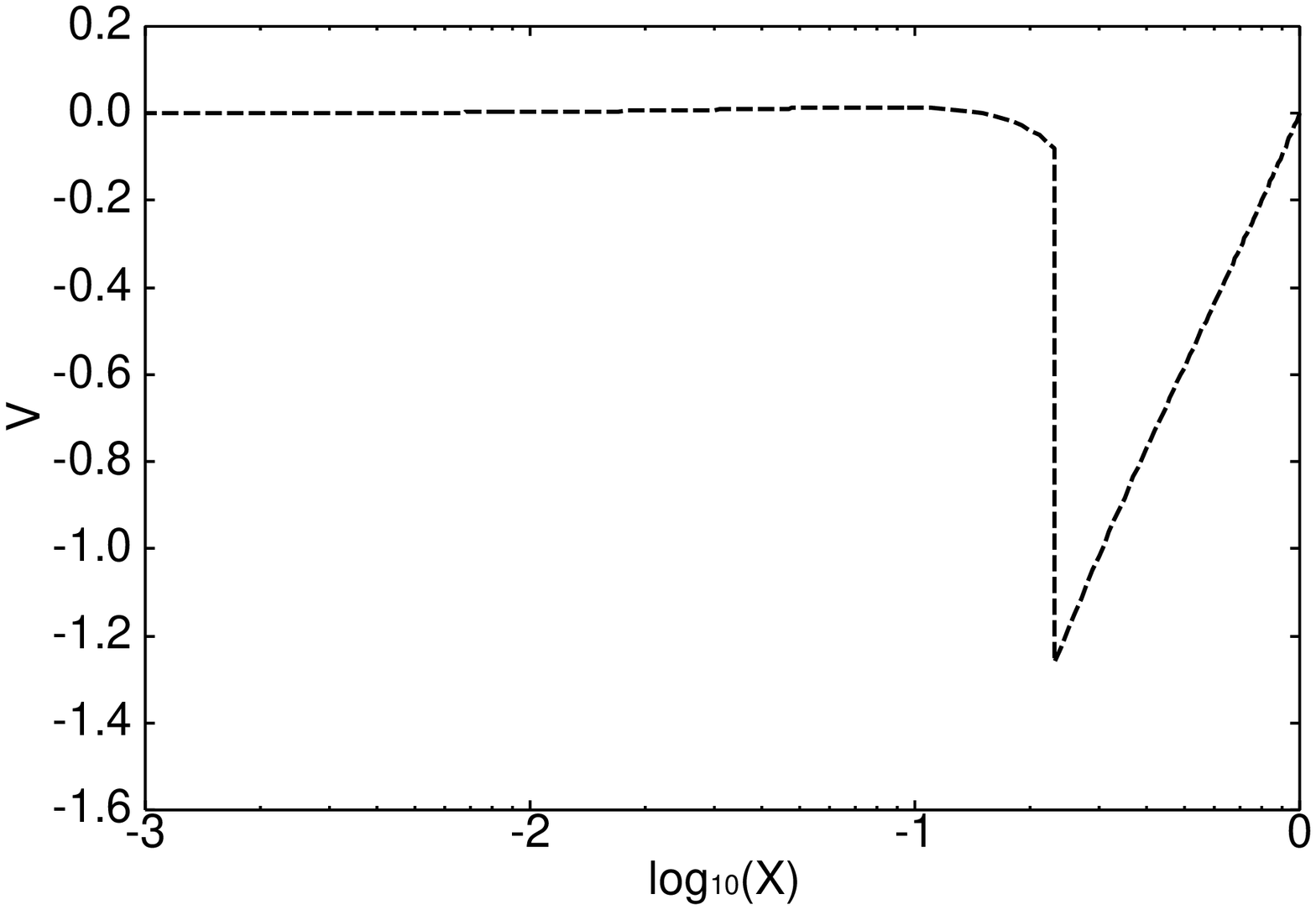}}\\
\resizebox{8cm}{!}{\includegraphics{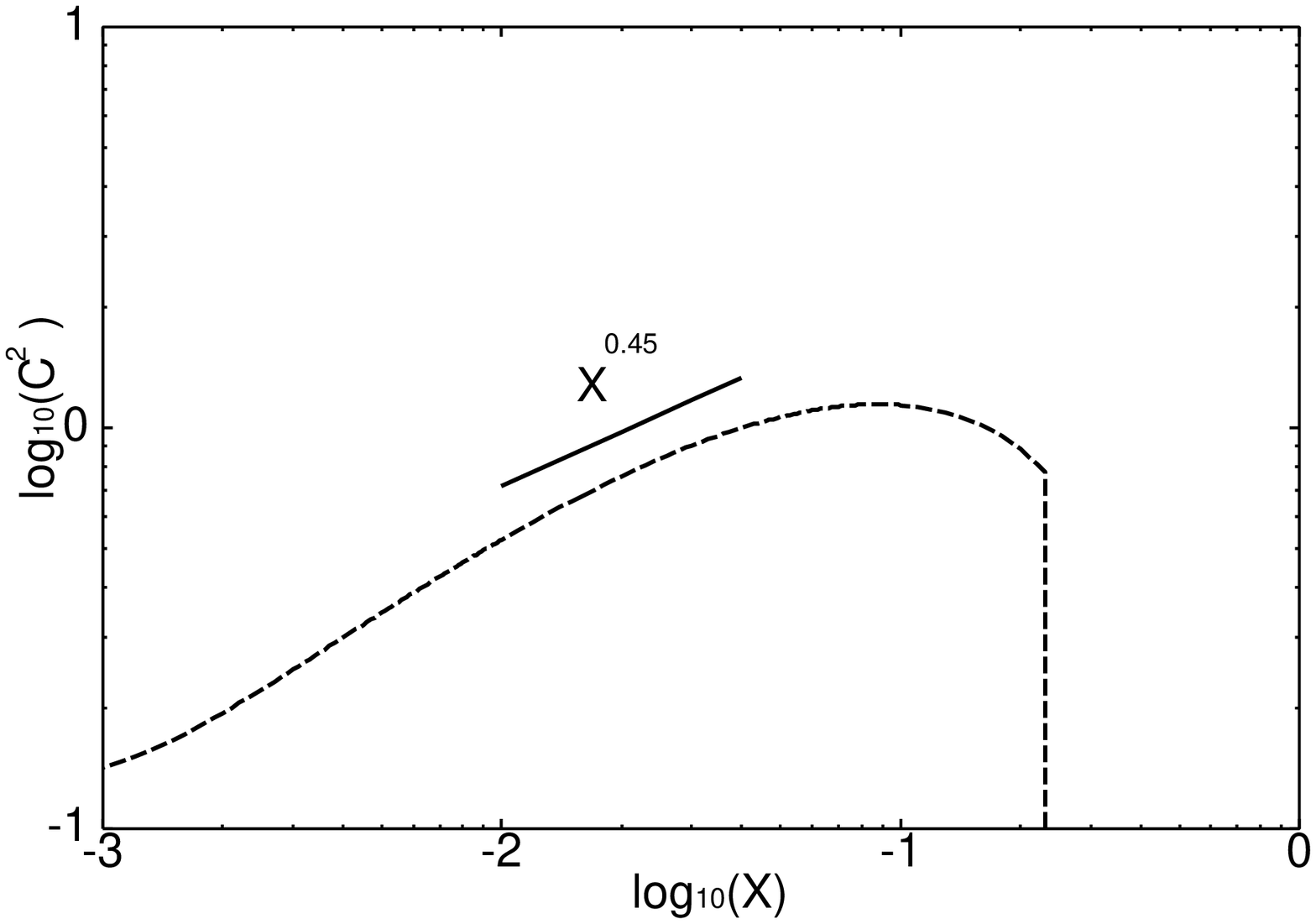}}%
\ \\
\caption{Similarity variables with $A$=2, $B$=5/2, and $K_{0}$=10$^{-4}$. 
The dashed line indicates the result with heating, ${\mathcal{H}}$=0.8. 
The panels indicate the same variable as in Fig. 6.}
\label{fig:figure8}
\end{figure}%
\begin{table}
 \centering
 \begin{center}
  \begin{tabular}{cccccc}\\
  \hline
   $\epsilon$ & ${\mathcal{H}}$ & $X_{s}$ & $u$ & $v$ & $w$ \\
  \hline
\hline
  \multicolumn{6}{c}{$K_{0}$=0.01}\\
\hline
     & 0   & 0.326595 & $-$2.32 & $-$0.39 & $+$0.63 \\
  1.0& 0.4 & 0.378087 & $-$2.13 & $-$0.29 & $-$0.70 \\
     & 0.8 & 0.438560 & $-$1.93 & $-$0.28 & $-$0.61 \\
  \hline
  \hline
  \multicolumn{6}{c}{$K_{0}$=0.1}\\
  \hline
     & 0   & 0.283055 & $-$2.48 & $-$0.52 & $+$0.91 \\
  1.0& 0.4 & 0.304111 & $-$2.36 & $-$0.47 & $+$0.59 \\
     & 0.8 & 0.323845 & $-$2.06 & $-$0.51 & $-$0.06 \\
  \hline
  \hline
\end{tabular}
\caption{Same as Table 2, except $A$=2 and $B$=$-$7/2.}
\label{tab:Table4}
\end{center}
\end{table}%
  
\begin{table}
 \centering
 \begin{center}
  \begin{tabular}{cccccc}\\
  \hline
   $K_{0}$ & ${\mathcal{H}}$ & $X_{s}$ & $u$ & $v$ & $w$ \\
  \hline
  \hline
  10$^{-4}$ & 0.7 & 0.224706 & $-$1.23 & $+$0.42 & $+$0.80 \\
            & 0.8 & 0.231401 & $-$1.10 & $+$0.45 & $+$0.82 \\
  \hline
  \hline
  10$^{-5}$ & 0.5 & 0.214030 & $-$1.32 & $+$0.36 & $+$0.52 \\
            & 0.8 & 0.232478 & $-$0.91 & $+$0.33 & $+$0.20 \\
  \hline
  \hline
  10$^{-6}$ & 0.3 & 0.203330 & $-$1.49 & $+$0.28 & $+$6.24 \\  
            & 0.8 & 0.232689 & $-$0.96 & $+$0.39 & $+$1.18 \\  
  \hline
  \hline
\end{tabular}
\caption{Properties of solutions for $\epsilon$=1/3($n$=$-$1), $A$=2, and 
$B$=5/2. The parameters are the same as those given in Table 2.}
\label{tab:Table5}
\end{center}
\end{table}

We consider the similarity solutions for the case that cooling function has a 
collisional process, $A$=2. We exclude the case of $\epsilon$=2/3 
because of $A$=3/2 from equation (\ref{eq:equation12}). Thus, we derive the 
similarity solutions with $\epsilon$=1 and 1/3. 
For $\epsilon$=1, $B$=$-$7/2. 
Table 4 gives the results of the slopes of the similarity variables. 
The density profiles are steeper than those of 
$\Lambda$$\propto$$\rho^{3/2}T$. 
Substituting ($A$,$B$,$w$)=(2,$-$7/2,0.91), which is the case of 
$K_{0}$=0.1 and ${\mathcal{H}}$=0, for equation (\ref{eq:equation46-2}), 
the density slope becomes $-$4.59, and the temperature slope becomes $-$1 
from equation (\ref{eq:equation46-3}). 
However, because $B$=$-$7/2, the energy loss is not 
large enough to produce a large infall velocity. As a result, 
the gradients of the density and temperature become about $-$2.5 and $-$0.5 
in the case of $K_{0}$=0.1. 

For $\epsilon$=1/3($n$=$-$1), the cooling function is 
$\Lambda$$\propto$ $\rho^{2}T^{5/2}$, in which case we only find 
the 'stagnation' solution in the cases of $K_{0}$=0.01 or 0.1. 
The reason is that the effect of cooling is very strong at the central 
region, leading to a cooling catastrophe. 
Therefore, we try to find heating solutions that can balance the cooling. 
As a result, when $K_{0}$=10$^{-4}$, 
we find that the heating solutions exist with 
${\mathcal{H}}$=0.7-0.8, and that the heating is so weak  
that a cooling catastrophe is unavoidable in the case of ${\mathcal{H}}<$0.7. 
Similarly, when $K_{0}$=10$^{-5}$, an 'eigensolution' exists with 
${\mathcal{H}}$=0.5-0.8. When $K_{0}$=10$^{-6}$, we find an 
'eigensolution' with ${\mathcal{H}}$=0.3-0.8. 
Table 5 gives its properties. When $K_{0}>$10$^{-4}$ and 
${\mathcal{H}}>$0.8, the heating is strong in the outer region where 
cooling can be neglected, and the expanding velocity becomes large 
and an 'eigensolution' is not found. 
Figure 8 shows the result of $\epsilon$=1/3, $K_{0}$=10$^{-4}$, 
and ${\mathcal{H}}$=0.8. Because the effect of heating is strong in 
the outer region, the slope of the density becomes shallow and 
the temperature decreases with decreasing radius, 
which are the same reasons as described 
for the previous model discussed in Section 3.2. 

\section{Discussion and Summary}
In this paper, we present self-similar solutions including cooling 
and heating for collisional gas. We assume that the cooling rate 
has a power-law dependences of the gas density and temperature, 
$\Lambda$$\propto$$\rho^{A}T^{B}$, and that the heating rate is 
$\Gamma$$\propto$$\rho T$. In order to obtain a similarity solution, 
$A$ and $B$ are selected by requiring that the cooling time is 
proportional to the dynamical time. 
The main results in this paper are the following:

\begin{enumerate}
\item In the region where the cooling rate is stronger than the heating rate, 
a cooling inflow is established, and the gas is compressed and heats up. 
Because the compression is stronger in the inner region than in the outer 
region, 
the temperature increases with decreasing radius. Furthermore, 
if the large infall velocity, $V$$<<$$-\xi X$, is produced due to an 
enormous energy loss, the temperature slope approaches $-$1, and the 
density slope approaches a value that depends on $A$, $B$, and the 
velocity slope $w$.
 
\item In the region where the heating rate is larger than the cooling rate, 
the infall velocity is suppressed by heating, the compression is weakened 
and the gas cools down. 
The slope of the density becomes shallow due to suppression of the collapse, 
and the temperature is lower than that without heating.  

\item For $\epsilon$=1/3 ($n$=$-$1) and ($A$,$B$)=(2,5/2), 
the 'stagnation' solutions are only derived in the cases of 
$K_{0}$=0.1 or 0.01. The reason for this is that the cooling 
is very strong in the inner region, and a cooling catastrophe occurs. 
We include the heating and a small $K_{0}$ in order to decrease 
the effect of cooling, and find heating solutions that can balance 
the cooling with $K_{0}$=10$^{-4}$ and ${\mathcal{H}}$=0.7-0.8.

\end{enumerate}

We consider that the self-similarity solution presented here can be used 
for modeling of structure formations.
For example, the value of the slope parameter, $\beta_{fit}$, 
obtained from observations of the surface brightness profile 
indicates the typical value for rich clusters, $\beta_{fit} \sim 2/3$. 
On the other hand, group systems have 
significantly flatter slopes with $\beta_{fit} \sim 0.4$
(Arnaud \& Evrard 1999; Helsdon \& Ponman 2000). 
As an origin of such discrepancies, it has been proposed that 
the intracluster medium (ICM) extends to the outer regions by 
non-gravitational heating, leading to flatter density profiles 
(Metzler \& Evrard 1997). 
Furthermore, in the model $A$=2 and $\epsilon$=1/3($n$=$-$1), 
which is close to those 
of real clusters, a cooling catastrophe occurs if there is no heating,
and there are the heating solutions that can balance the cooling. 
This may mean that real clusters oscillate between a heating dominated regime 
and one with a cooling catastrophe (Kaiser \& Binney 2003), or 
that the heating sources, such as AGN (Churazov et al. 2001; 
Quilis et al. 2001; Reynolds et al. 2002; Br$\ddot{\rm{u}}$ggen et al. 2002; 
Br$\ddot{\rm{u}}$ggen \& Kaiser 2002; Br$\ddot{\rm{u}}$ggen 2003) 
or thermal conduction 
(Fabian et al. 2002; Ruszkowski \& Begelman 2002; 
Voigt et al. 2002; Zakamska \& Narayan 2003), balance the cooling.
Of course, it is difficult to directly apply the solution presented here 
to a real cluster, because the time evolutions of the cooling and 
heating differ from the Hubble time. For a realistic treatment of 
this problem, it is necessary to perform a numerical simulation while 
assuming regular radiative cooling and heating rates. 
However, the accuracy of the numerical simulation is limited by 
the finite resolution. We consider that a comparison between the 
simulation and the similarity solution 
gives insights into the thermal evolution of clusters of galaxies.

\section*{Acknowledgments}

We would like to thank an anonymous referee for very useful comments and 
suggestions, which have improved the quality of our work.

\bsp

\label{lastpage}

\end{document}